\documentclass[aps,pra,amsmath,amssymb,twocolumn,superscriptaddress]{revtex4-1}
\usepackage{ulem}
\usepackage{amsmath,amssymb}
\usepackage{graphicx}	
\usepackage{xcolor}

\def\vecb{\boldsymbol}

\def\dc{\mathrm{dc}}
\def\rec{\mathrm{rec}}
\def\sft{\mathrm{sft}}
\def\inj{\mathrm{inj}}

\begin{document}

\author{Shunsuke~A.~Sato}
\email{ssato@ccs.tsukuba.ac.jp}
\affiliation 
{Center for Computational Sciences, University of Tsukuba, Tsukuba 305-8577, Japan}
\affiliation 
{Max Planck Institute for the Structure and Dynamics of Matter, Luruper Chaussee 149, 22761 Hamburg, Germany}

\author{Angel~Rubio}
\email{angel.rubio@mpsd.mpg.de}
\affiliation 
{Max Planck Institute for the Structure and Dynamics of Matter, Luruper Chaussee 149, 22761 Hamburg, Germany}
\affiliation 
{Center for Computational Quantum Physics, Flatiron Institute, 162 Fifth Avenue, New York, NY 10010, USA}

\title{Limitations of mean-field approximations in describing shift-current and injection-current in materials}

\begin{abstract}
We theoretically investigate bulk photovoltaic effects, with a specific focus on shift-current and injection-current. Initially, we perform a numerical analysis of the direct current (dc) induced by a laser pulse with a one-dimensional model, utilizing mean-field theories such as time-dependent Hartree--Fock and time-dependent Hartree methods. Our numerical results, obtained with mean-field theories, reveal that the dc component of the current exists even after irradiation with linearly polarized light as a second-order nonlinear effect, indicating the generation of injection current. Conversely, when we employ the independent particle approximation, no injection current is generated by linearly polarized light. To develop the microscopic understanding of injection current within the mean-field approximation, we further analyze the dc component of the current with the perturbation theory, employing the mean-field approximations, the independent-particle approximation, and the exact solution of the many-body Schr\"odinger equation. The perturbation analysis clarifies that the injection current induced by linearly polarized light under the mean-field approximations is an artifact caused by population imbalance, created through quantum interference from unphysical self-excitation pathways. Therefore, investigation of many-body effects on the bulk photovoltaic effects have to be carefully conducted in mean-field schemes due to potential contamination by unphysical dc current. Additionally, we perform the first-principles electron dynamics calculation for BaTiO$_3$ based on the time-dependent density functional theory, and we confirm that the above findings from the one-dimensional model calculation and the perturbation analysis apply to realistic systems.
\end{abstract}

\maketitle
%
%
%
%
\section{Introduction \label{sec:intro}}

The bulk photovoltaic effect, the conversion of light into an electric current, underpins technological applications within modern society. Among the various mechanisms associated with the photovoltaic effect, the bulk photovoltaic effect has been intensively studied from both a fundamental and technological points of view \cite{PhysRevLett.107.126805,doi:10.1126/science.1168636,Yang2010}. This effect, formerly referred to as the anomalous photovoltaic effect in inversion symmetry broken materials \cite{KOCH1975847,doi:10.1080/00150197408234118,doi:10.1080/00150197608236593,Kratzig_1977,doi:10.1080/713819567,doi:10.1080/00150197608236596} and nowadays identified as the shift current \cite{PhysRevB.61.5337}, is a second-order nonlinear optical effect that generates a direct current (dc) in the presence of light.

Historically, theoretical examinations of the shift current encountered a divergence problem in the nonlinear susceptibility at the low-frequency limit. This issue was addressed by developing an explicit expression for the nonlinear susceptibility, using perturbation theory within the framework of the independent particle approximation \cite{PhysRevB.61.5337}. This expression, once coupled with first-principles electronic structure calculations, has been used extensively to analyze the shift current in real materials \cite{PhysRevLett.109.116601,PhysRevLett.109.236601,Cook2017,PhysRevB.97.241118,PhysRevB.101.045104}. Concurrently, experimental research on the shift current has broadened, offering a variety of insights \cite{10.1063/1.2131191,PhysRevB.96.241203,10.1063/1.5055692,Nakamura2017,doi:10.1073/pnas.1802427116,10.1063/1.5087960}.

Despite the considerable interest in these phenomena, the majority of theoretical investigations into shift currents in actual materials have been limited to the independent particle approximation. Recently, there have been several attempts to elucidate the role of many-body effects on the microscopic mechanisms of shift current. For example, Chan \textit{et al} suggested substantial enhancement of the shift current due to excitonic effects, utilizing the GW plus Bethe-Salpeter approach \cite{doi:10.1073/pnas.1906938118}. Furthermore, Kaneko \textit{et al} proposed an enhancement of the photovoltaic effect in excitonic insulators via collective excitations \cite{PhysRevLett.127.127402}. These studies indicate that many-body effects could have a pivotal role in enhancing photovoltaic effects. Nevertheless, the understanding of many-body effects in the microscopic mechanisms of the photovoltaic effects is still incomplete. For further development of detailed understanding of the many-body effects, it is crucial to assess the applicability of different theoretical approximations to accurately interpret these phenomena.

In this study, we conduct a numerical investigation of the bulk photovoltaic effects with a one-dimensional model, employing mean-field theories such as the time-dependent Hartree--Fock (TDHF) and time-dependent Hartree (TDH) methods, to assess the suitability of mean-field approximations. We also analyze the bulk photovoltaic effects through perturbation theory utilizing the independent particle approximation, mean-field approximation, and the exact many-body Schr\"odinger equation. Our results indicate that the photo-induced current under the mean-field approximation displays qualitatively different behaviors compared to both the independent particle approximation and the exact Schr\"odinger equation. Specifically, the mean-field approximation generates the dc current as the second-order nonlinear effect even after irradiation of linearly polarized light, which neither the independent particle approximation nor the exact Schrodinger equation can describe. This suggests that the mean-field approximation could artificially induce the injection current under linearly polarized light. Consequently, analyses of the bulk photovoltaic effect using a mean-field approximation could significantly overestimate the effect due to this artifact, as the divergence of the susceptibility tensor of the injection current could completely overcome the susceptibility of the intrinsic shift current. Additionally, we perform the electron dynamics calculations for BaTiO$_3$ based on time-dependent density functional theory (TDDFT) to confirm the findings from the one-dimensional model and the perturbation analysis apply to realistic systems.

The paper is organized as follows. In Sec.~\ref{sec:method}, we describe theoretical and numerical methods for investigating the bulk-photovoltaic effect. In Sec.~\ref{sec:results}, we show the numerical results obtained by the method introduced in Sec.~\ref{sec:method} and discuss the qualitative difference among the results obtained with different approaches. In Sec.~\ref{sec:perturbation}, we analyze the bulk photovoltaic effect based on perturbation theory and explore the microscopic origin of the qualitative difference among different approximations. In Sec.~\ref{sec:tddft}, we perform the first-principles electron dynamics calculation to analyze the shift-current and confirm the findings from the previous sessions for a realistic material. Finally, our findings are summarized in Sec.~\ref{sec:summary}.

\section{Methods \label{sec:method}}

In this study, the one-dimensional TDHF method is mainly utilized to simulate light-induced electron dynamics in solids. Each electronic orbital consisting a Slater determinant obeys the following TDHF equation,
\begin{widetext}
\begin{align}
i\hbar \frac{\partial}{\partial t}u_{b, k_x}(x,t)=
\left [
\frac{1}{2m_e}\left(-i\hbar\frac{\partial}{\partial x}+\hbar k_x + eA_x(t) \right )^2 + v_{\mathrm{ion}}(x) + v_{H}(x,t) + \hat v_F(t)
\right ]u_{b, k_x}(x,t),
\label{eq:tdhf}
\end{align}
\end{widetext}
where $b$ is the band index, $k_x$ is the Bloch wavenumber, and $u_{b,k_x}(x,t)$ represents the periodic part of the Bloch wavefunction which satisfies $u_{b,k_x}(x+a,t)=u_{b,k_x}(x,t)$, with the lattice constant, $a$. Here, $A_x(t)$ is a homogeneous vector potential related to an external electric field as $E_x(t)=-dA_x(t)/dt$. The ionic potential is denoted as $v_{\mathrm{ion}}(x)$, and the spatial periodicity is imposed as $v_{\mathrm{ion}}(x+a)=v_{\mathrm{ion}}(x)$. In this study, the ionic potential, $v_{\mathrm{ion}}(x)$, is defined as
\begin{align}
v_{\mathrm{ion}}(x) = \int^{\infty}_{-\infty}dx' w(x-x')\rho_{\mathrm{ion}}(x'),
\end{align}
where $\rho_{\mathrm{ion}}(x)$ is the ionic charge density. Here, $w(x)$ denotes the one-dimensional soft-Coulomb interaction and is defined as
\begin{align}
w(x) = \beta \frac{e^2}{\sqrt{x^2+a^2_0}},
\end{align}
where $\beta$ is a dimensionless adjustable parameter, and $a_0$ is the Bohr radius. Therefore, for this study, the softening parameter of the soft-Coulomb potential is set to $a_0$. To introduce the standard Hartree potential $v_{H}(x,t)$ and the Fock operator $\hat v_F(t)$, the one-body reduced density matrix and the one-body density are defined as
\begin{align}
\rho_{DM}(x,x',t)&=\frac{2}{N_k}
\sum_{b, k_x} u_{b, k_x}(x,t)u^*_{b, k_x}(x',t) \nonumber \\
& \times \exp \left [ i \left ( k_x+\frac{eA(t)}{\hbar} \right) \left (x-x' \right) \right ], \label{eq:one-body-dm}\\
\rho(x,t) &= \rho_{DM}(x,x,t), \label{eq:one-body-density}
\end{align}
where $N_k$ is the number of $k$-points in the calculation. Using the one-body density, $\rho(x,t)$, the Hartree potential is defined as
\begin{align}
v_H(x,t)=\int^{\infty}_{-\infty}dx' w(x-x') \rho(x,t).
\end{align}

Furthermore, the Fock operator is defined using the one-body density matrix as follows:
\begin{widetext}
\begin{align}
\hat v_F(t) u_{b,k_x}(x,t) = -\frac{1}{2}
\int^{\infty}_{-\infty}dx'w(x-x') \rho_{DM}(x,x',t)
\exp \left [ -i \left ( k_x+\frac{eA(t)}{\hbar} \right) \left ( x - x'\right ) \right ]
u_{b, k_x}(x',t).
\end{align}
\end{widetext}

By solving the TDHF equation, Eq.~(\ref{eq:tdhf}), we can elucidate the electron dynamics induced by the vector potential $A_x(t)$. Furthermore, by utilizing the time-dependent electron orbitals, $u_{b,k_x}(x,t)$, we can evaluate physical quantities within the time domain. For instance, the induced electric current can be calculated as follows:
\begin{align}
J(t)=-\frac{2e}{aN_k} \sum_{b,k_x}\int^a_0dx u^*_{b,k_x}(x,t)
\hat v_x(t)
u_{b,k_x}(x,t),
\label{eq:current}
\end{align}
where the velocity operator $\hat v_x(t)$ is defined as:
\begin{align}
\hat v_x (t) = \frac{1}{m_e}\left [-i\hbar \frac{\partial}{\partial x}+\hbar k_x + eA_x(t) \right ].
\end{align}

To model a solid-state system with broken inversion symmetry, we employ the following ionic charge distribution, denoted as $\rho_{\mathrm{ion}}(x)$:
\begin{align}
\rho_{\mathrm{ion}}(x) = \sum_{n=-\infty}^{\infty} \left[
-3 \delta(x+n\cdot a)- \delta \left (x+\frac{a}{3}+n\cdot a \right )
\right ].
\end{align}

In this study, the lattice constant $a$ is set to $4.0352$~$\AA$, analogous to that of BaTiO$_3$~\cite{Kwei1993}, a material typically utilized for investigating shift-current. To impose charge neutrality, we incorporate four electrons within the unit cell for this study. Furthermore, we set the dimensionless parameter $\beta$ of the soft-Coulomb interaction to $0.2915$~a.u. so as to reproduce the band gap of BaTiO$_3$ with the model calculation.

\begin{figure}[ht]
\includegraphics[width=0.97\linewidth]{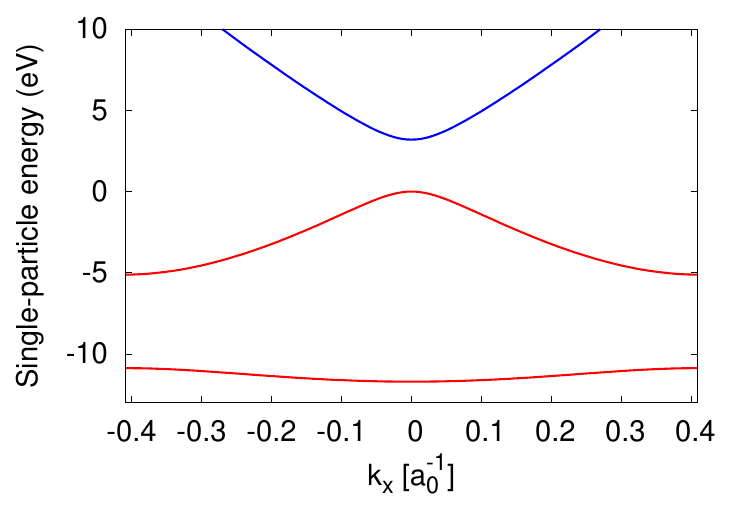}
\caption{\label{fig:bandstructure}
Computed bandstructure of the one-dimensional model of a bulk with broken inversion symmetry with the static Hartree--Fock method. The two spin-degenerate occupied bands are described as the red solid line, whereas the conduction bands are described as the blue solid line.
}
\end{figure}

Figure~\ref{fig:bandstructure} shows the band structure obtained by solving the static Hartree--Fock equation under the conditions stated above. The red solid line denotes the two filled bands (valence bands), while the blue solid line illustrates the empty band (conduction bands). The calculated band-gap at the $\Gamma$-point ($k_x=0$) is 3.2~eV.

For the subsequent calculations in Section~\ref{sec:results}, the real-space coordinate $x$ within the unit cell ($0\le x \le a$) is discretized into $16$ grid points. Similarly, the First Brillouin zone is divided into $1025$ $k$-points.

\section{Results}\label{sec:results}

\subsection{Linear Optical Properties}\label{subsec:linear}

To develop insights into optical responses of solids, we first revisit linear optical properties of the system with mean-field approximations. For this purpose, we calculate the electron dynamics in the presence of an impulsive distortion given by
\begin{equation}
A_x(t) = A_0 \Theta(t),
\label{eq:impulse}
\end{equation}
where $\Theta(t)$ is the Heaviside step function, and $A_0$ is the amplitude of the distortion. In this study, we set $A_0$ to $10^{-4}$~a.u. so that the induced current is proportionate to the amplitude $A_0$, resulting in the linear response.

The TDHF equation, Eq.~(\ref{eq:tdhf}), is solved with the vector potential as defined in Eq.~(\ref{eq:impulse}). The static Hartree--Fock method is utilized to compute the ground state of the system, which is subsequently used as the initial condition for Eq.~(\ref{eq:tdhf}). The induced current, $J(t)$, is computed using Eq.~(\ref{eq:current}).

Assuming that the amplitude $A_0$ is sufficiently small, the optical conductivity $\sigma(\omega)$ of the system can be evaluated as
\begin{equation}
\sigma(\omega)=-\frac{1}{A_0}\int^{T_{\mathrm{sim}}}_0 dt
J(t)e^{i\omega t} W\left (\frac{t}{T_{\mathrm{sim}}} \right ),
\end{equation}
where $T_{\mathrm{sim}}$ represents the simulation time period, and $W(x)$ is a window function introduced to suppress numerical noise in the Fourier transform resulting from finite simulation time. For practical calculations to analyze the linear responses, the simulation time, $T_{\mathrm{sim}}$, is set to $20$~fs. Furthermore, we employ the following form of the window function in this study:
\begin{equation}
W(x) = \cos^4 \left ( \frac{\pi}{2}x\right ).
\end{equation}

\begin{figure}[ht]
\includegraphics[width=0.97\linewidth]{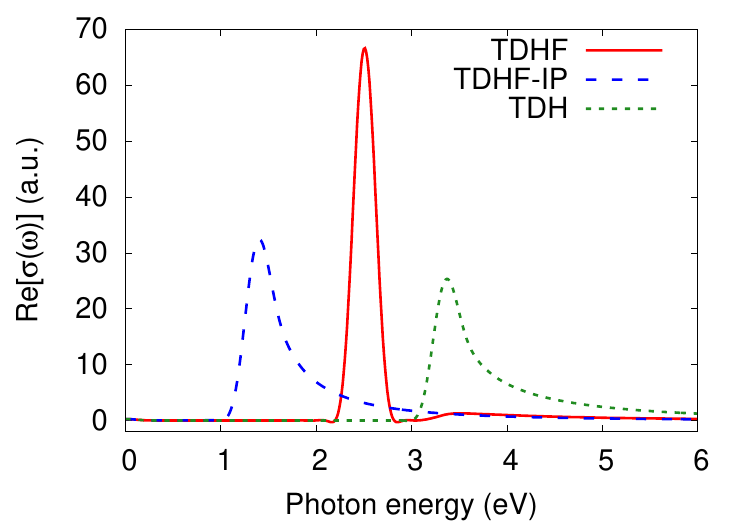}
\caption{\label{fig:sigma_w}
Real-part of conductivity $\sigma(\omega)$ of the one-dimensional solid-state model. The result obtained using the TDHF method is described as red solid line, the result obtained using the independent particle approximation is described as blue dashed line, and the result obtained using the TDH method is described as green dotted line.}
\end{figure}

Figure~\ref{fig:sigma_w} shows the real-part of the conductivity of the one-dimensional solid-state system, computed with the parameters described in Sec.~\ref{sec:method}. The red solid line represents the results obtained by using the TDHF method, revealing a distinctive peak structure below the band gap of 3.2~eV. This peak can be understood as the excitation to an excitonic state.

To obtain insights into the characteristics of the mean-field approximation within the context of the linear optical response, we introduce two distinct approximations to the TDHF method. The first approximation is the independent-particle approximation, denoted as TDHF-IP, which freezes the time-dependence of the Hartree potential and Fock operator in the TDHF method. The second approximation is the time-dependent Hartree approximation, denoted as TDH.

When employing the TDHF-IP approach, the static Hartree-Fock method is employed to prepare wavefunctions and eigenvalues as the initial conditions for the time propagation. In this approximation, we neglect the time dependencies of $v_H(x,t)$ and $\hat v_F(t)$ by replacing the time-dependent one-body density matrix, $\rho_{DM}(x,x',t)$, with $\rho_{DM}(x,x',0)$ for the computation of $v_H(x,t)$ and $\hat v_F(t)$. Therefore, within the TDHF-IP approximation, we disregard the time dependencies of the mean-field potential completely. The resulting dynamics can be interpreted as those of an independent particle system under a corresponding external potential.

By contrast, the TDH method removes the Fock operator, $\hat v_F(t)$, from the TDHF process during both the preparation of the ground state and the time propagation of the system. Consequently, the mean-field potential consists solely of the Hartree potential, $v_H(x,t)$. This approximation provides one of the simplest mean-field approximations for quantum many-body systems.

In Fig.~\ref{fig:sigma_w}, the blue dashed line represents the result obtained from the TDHF-IP method, while the green dotted line denotes the result obtained from the TDH method. The TDHF-IP result indicates the vanishing of the excitonic peak below the gap, with the spectral weight of the excitonic peak absorbed by the above-gap absorption, thus reflecting a nature of independent particle systems – the absence of excitons. The TDH method yields a spectral structure similar to that of TDHF-IP but with a substantial red-shift. Such similar spectral structures imply that the TDH method does not capture the excitonic contribution to the excitation spectrum. The red-shift of the spectrum in the TDH method reflects the band gap reduction caused by the exclusion of the exchange interaction. These insights confirm the widely accepted understanding in condensed matter physics: the static contribution of the Fock operator enlarges the gap, while the dynamical contribution of the Fock operator describes the excitonic response of the lowest optical transition~\cite{RevModPhys.74.601}. Employing these three methods, we further analyze nonlinear optical responses in the subsequent section.

\subsection{Second-Order Nonlinear Optical Responses: Shift Current and Injection Current \label{subsec:second}}

Having revisited the mean-field characteristics within linear optical responses, we extend our investigation to nonlinear optical phenomena, with a specific focus on the second-order nonlinear optical effects, namely shift current and injection current~\cite{PhysRevB.61.5337}. To obtain insights into these second-order nonlinear optical phenomena, we first review the nonlinear susceptibility in the frequency domain, and the corresponding current dynamics in the time domain.

In this analysis, we consider a one-dimensional case to maintain simplicity, though the approach can be straightforwardly generalized to two or three dimensional systems. The second-order nonlinear polarization $P^{(2)}(t)$, induced by an electric field $E(t)$, can be described as \cite{boyd2020nonlinear}:
\begin{align}
P^{(2)}(t) = \int^{\infty}_{-\infty}dt' \int^{\infty}_{-\infty}dt'' \chi^{(2)}(t-t',t-t'')E(t')E(t''),
\label{eq:second-pol-time}
\end{align}
where $\chi^{(2)}(t-t',t-t'')$ represents the second-order nonlinear susceptibility in the time domain. By applying the Fourier transform to Eq.~(\ref{eq:second-pol-time}), we derive the following relation:
\begin{align}
&\tilde P^{(2)}(\omega_{\Sigma}) =\frac{1}{2\pi} \int^{\infty}_{-\infty}d\omega' \int^{\infty}_{-\infty}d\omega'' \delta (\omega_{\Sigma}-\omega'-\omega'') \nonumber \\
&\times \tilde \chi^{(2)}(\omega_{\Sigma}; \omega', \omega'') \tilde E(\omega') \tilde E(\omega'') \nonumber \\
&=\frac{1}{2\pi} \int^{\infty}_{-\infty}d\omega' \tilde \chi^{(2)}(\omega_{\Sigma}; \omega', \omega_{\Sigma}-\omega') \tilde E(\omega') \tilde E(\omega_{\Sigma}-\omega'),
\label{eq:second-pol-freq}
\end{align}
where $\tilde P^{(2)}(\omega)$, $\tilde \chi^{(2)}(\omega; \omega', \omega'')$, and $\tilde E(\omega)$ are the Fourier transformations of $P^{(2)}(t)$, $\chi^{(2)}(t-t',t-t'')$, and $E(t)$, respectively.

Optical rectification, shift current, and injection current constitute second-order nonlinear dc optical responses. These phenomena can be characterized based on the divergent behavior of the second-order nonlinear susceptibility, $\tilde \chi(\omega_{\Sigma},\omega',\omega'')$. In the low frequency limit $(\omega_{\Sigma}=\omega'+\omega'' \rightarrow 0)$, the nonlinear susceptibility can be described as \cite{PhysRevB.61.5337}:
\begin{align}
\tilde \chi(\omega_{\Sigma},\omega',\omega'')=\tilde \chi^{(2)}_{\rec}(\omega', \omega'')
+\frac{\tilde \sigma^{(2)}_\sft (\omega', \omega'')}{-i\omega_{\Sigma}}
+\frac{\tilde \eta^{(2)}_\inj (\omega', \omega'')}{(-i\omega_{\Sigma})^2},
\label{eq:second-opt-nonlinear-divergence}
\end{align}
where $\tilde \chi^{(2)}_{\rec}(\omega', \omega'')$, $\tilde \sigma^{(2)}_\sft(\omega', \omega'')$, and $\tilde \eta^{(2)}_\inj(\omega', \omega'')$ denote regular analytic functions, corresponding to optical rectification, shift current, and injection current, respectively.

For a deeper understanding of these nonlinear optical phenomena, a time-domain behavior of induced responses complements their divergent behavior of the susceptibilities in the frequency domain. We conduct this exploration by analyzing the dynamics induced by a laser pulse represented as:
\begin{align}
E(t)=f(t)\cos(\omega_0 t),
\label{eq:long-pulse}
\end{align}
where, $f(t)$ is the envelope function of the laser pulse, and $\omega_0$ is the average frequency. For this analysis, we assume that the envelope function $f(t)$ exhibits slow temporal variation, and that the Fourier transform of Eq.~(\ref{eq:long-pulse}) is concentrated predominantly around $\omega=\omega_0$.

We initiate our analysis with the polarization, $P^{(2)}_{\rec}(t)$, associated with optical rectification. The polarization induced by the pulse of Eq.~(\ref{eq:long-pulse}) can be evaluated as follows:
\begin{align}
P^{(2)}_{\rec}(t)&=\frac{1}{2\pi}\int^{\infty}_{-\infty}d\omega \tilde P^{(2)}_{\rec}(\omega)e^{-i\omega t} \nonumber \\
&=\frac{1}{2\pi}\int^{\infty}_{-\infty}d\omega \tilde \chi^{(2)}_{\rec}(\omega', - \omega') \tilde E(\omega') \tilde E(\omega-\omega') e^{-i\omega t} \nonumber \\
&\approx \frac{1}{2\pi}\tilde \chi^{(2)}_{\rec}(\omega_0, - \omega_0) \int^{\infty}_{-\infty}d\omega  \tilde E(\omega') \tilde E(\omega-\omega') e^{-i\omega t} \nonumber \\
&=\chi^{(2)}_{\rec}(\omega_0, - \omega_0) E^2(t).
\label{eq:opt-rec-full}
\end{align}

Here, we adopted the assumption that the Fourier transform of $E(\omega)$ is predominantly localized around $\omega=\omega_0$. By isolating the low frequency component of Eq.~(\ref{eq:opt-rec-full}), we extract the dc-like component of $P^{(2)}_{\rec}(t)$ as
\begin{align}
P^{(2)}_{\rec,\dc}(t)=\chi^{(2)}_{\rec}(\omega_0, - \omega_0) f^2(t).
\label{eq:opt-rec-dc}
\end{align}

The results demonstrate that the polarization associated with optical rectification is directly proportional to the square of the envelope function, $f(t)$. Similar analysis can be performed for the current $J^{(2)}_{\sft,\dc}(t)$ associated with the shift current, and the acceleration $K^{(2)}_{\inj,\dc}(t)$ associated with the injection current. This investigation yields the following relation:
\begin{align}
J^{(2)}_{\sft,\dc}(t)&=\frac{d}{dt}P^{(2)}_{\sft,\dc}(t)=
\sigma^{(2)}_{\sft}(\omega_0, - \omega_0) f^2(t), \label{eq:sft-curr-dc} \\
K^{(2)}_{\inj,\dc}(t)&=\frac{d^2}{dt^2}P^{(2)}_{\inj,\dc}(t)=
\eta^{(2)}_{\inj}(\omega_0, - \omega_0) f^2(t). \label{eq:inj-curr-dc}
\end{align}

\begin{figure}[htb]
\includegraphics[width=0.90\linewidth]{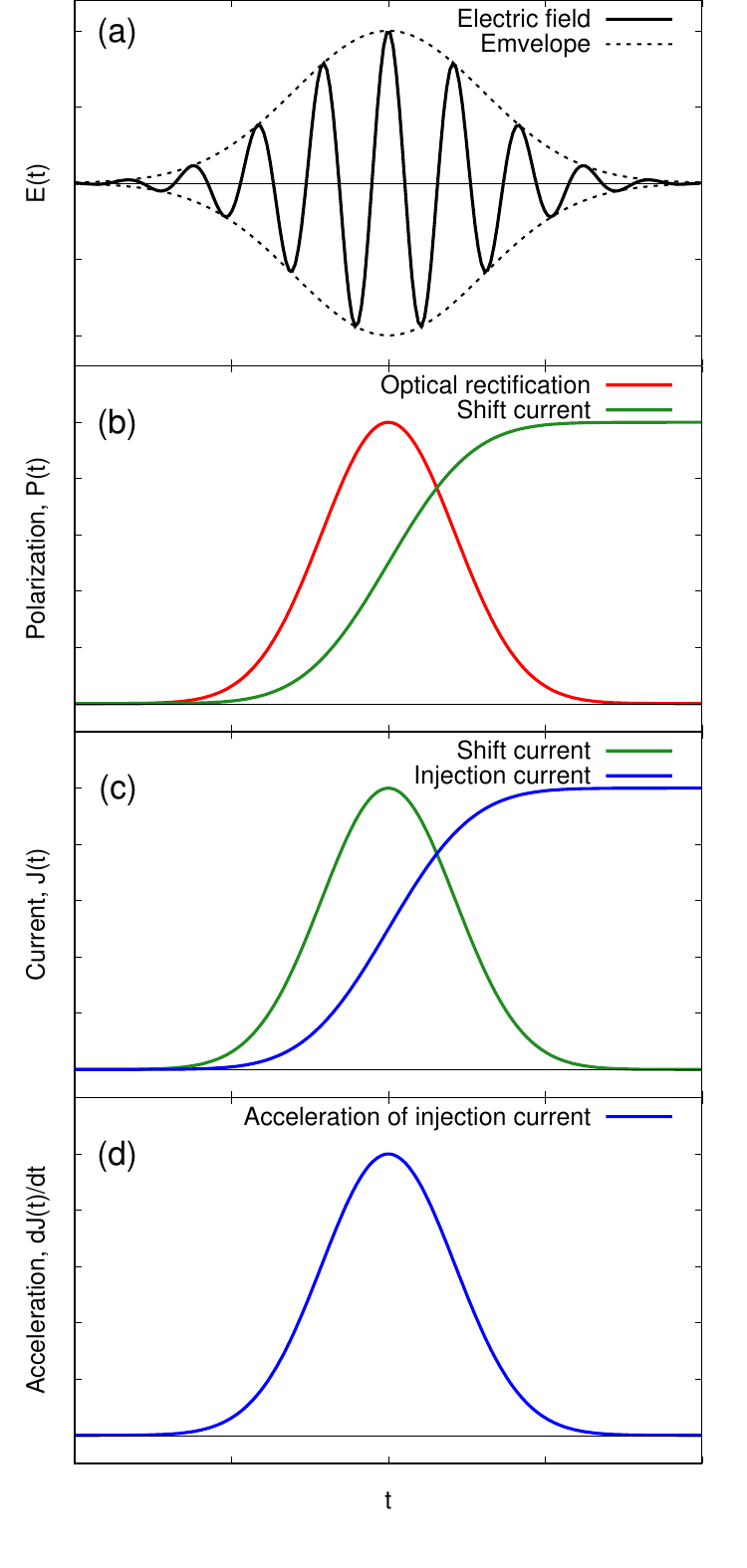}
\caption{\label{fig:chi2_res}
Schematics of the second-order nonlinear optical responses. (a) The time-profile of an applied example pulse is shown. (b) The polarizations associated with the optical rectification (red) and the shift current (green) are shown as a function of time. (c) The induced currents associated with the shift current (green) and the injection current (blue) are shown. (d) The acceleration associated with the injection current is shown.
}
\end{figure}

Figure~\ref{fig:chi2_res} shows a graphical representation of the time-domain behaviors of optical rectification, shift current, and injection current as described by Eqs.~(\ref{eq:opt-rec-dc}--\ref{eq:inj-curr-dc}). The time profile of a sample laser pulse is depicted in Fig.~\ref{fig:chi2_res}~(a), with the envelope function of the laser pulse represented by a black dashed line. The time profile of the polarization for optical rectification and shift current is shown in Fig.~\ref{fig:chi2_res}~(b). The polarization associated with optical rectification shifts only during laser irradiation, whereas the polarization associated with shift current remains finite even after the laser field ends. Figure~\ref{fig:chi2_res}~(c) shows the time profile of the currents associated with shift and injection currents. The shift current is only induced during laser irradiation, while the injection current remains finite even after the laser field ends. This behavior of the injection current can be understood by the fact that the acceleration, $K^{(2)}_{\inj,\dc}(t)$, associated with the injection current shifts only during laser irradiation, as displayed in Fig.~\ref{fig:chi2_res}~(d).

Guided by the time-domain behavior of the second-order nonlinear optical responses as shown in Fig.~\ref{fig:chi2_res}, we then numerically examine the impact of mean-field approximations on these nonlinear responses by employing the TDHF method. 

For practical calculations, we compute the electron dynamics induced by the following vector potential:
\begin{align}
A_x(t)=-\frac{E_0}{\omega_0}\cos \left (\omega_0 t + \phi_{\mathrm{CEP}} \right )
\cos^4 \left [\frac{\pi}{T_{\mathrm{pulse}}}t \right ]
\label{eq:laser-pulse-vec}
\end{align}
in the domain $-T_{\mathrm{pulse}}/2 \le t \le T_{\mathrm{pulse}}/2$, and zero outside this domain. Here, $E_0$ is the peak field strength, $\omega_0$ is the mean frequency, $\phi_{\mathrm{CEP}}$ is the carrier envelope phase (CEP), and $T_{\mathrm{pulse}}$ is the pulse duration. In this work, we set $E_0$ to $2\times 10^{-4}$~a.u., $\omega$ to $3.3$~eV/$\hbar$, and $T_{\mathrm{pulse}}$ to $40$~fs. We determined the field strength $E_0$ such that the resulting dc-like current is dominated by second-order nonlinear optical responses. Furthermore, we choose the photon energy $\hbar \omega_0$ to exceed the band gap, fulfilling the condition to induce the shift current \cite{PhysRevB.61.5337}.

We treat $\phi_{\mathrm{CEP}}$ as a tunable parameter to extract the dc-like component of the induced current. We denote the current induced by the laser field of Eq.~(\ref{eq:laser-pulse-vec}) as $J_x(x,\phi_{\mathrm{CEP}})$, explicitly noting $\phi_{\mathrm{CEP}}$ dependence. To extract dc-like component of the induced current, we consider the CEP average of the induced current as
\begin{align}
J_{x,\dc}(t) = \frac{1}{2\pi}\int^{2\pi}_0 d\phi_{\mathrm{CEP}} J_x(x,\phi_{\mathrm{CEP}}).
\label{eq:dc-current-phi-integral}
\end{align}

By utilizing this average, we effectively eliminate high-frequency components in the induced current, thereby extracting the dc-like component. For practical analysis, we calculate the integral of Eq.~(\ref{eq:dc-current-phi-integral}) as the mean of four values of $\phi_{\mathrm{CEP}}$: $\phi_{\mathrm{CEP}}=0,\pi/2,\pi,3\pi/2$.

\begin{figure}[htb]
\includegraphics[width=0.97\linewidth]{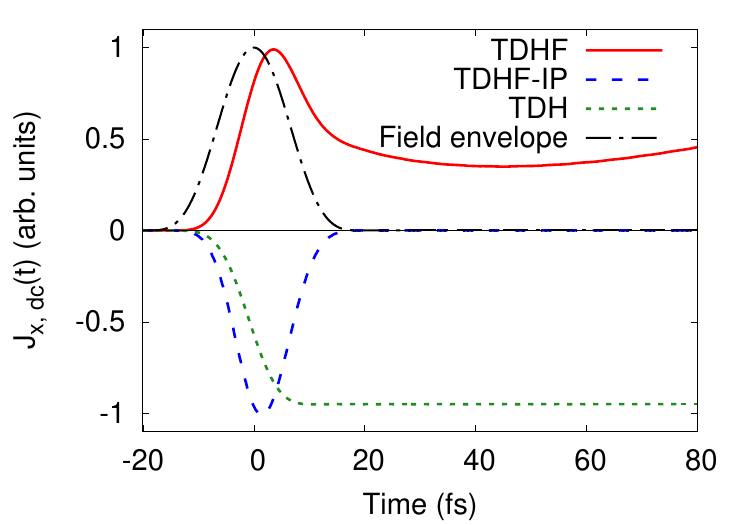}
\caption{\label{fig:current_t}
The slow component of the second order induced current, $J_{x,\dc}(t)$ is shown. The result obtained using the TDHF method is shown as the red solid line, and that obtained using the independent particle approximation is shown as the blue dashed line, and that obtained using the TDH method is shown as the green dotted line. As a guide, the envelope function of the applied electric field of Eq.~(\ref{eq:laser-pulse-vec}) is shown as black dash-dot line.
}
\end{figure}

Figure~\ref{fig:current_t} shows the time-evolution of the dc-like component of the induced current, $J_{x,\dc}(t)$, calculated using Eq.~(\ref{eq:dc-current-phi-integral}). As shown in Fig.~\ref{fig:current_t}, there are distinct differences in the behaviors of the results obtained by the TDHF, TDH, and TDHF-IP methods. Both TDHF and TDH methods show a finite current even after the laser field ends. By contrast, the TDHF-IP method does not present any residual current. These observations suggest that mean-field theories, such as TDHF and TDH, produce qualitatively distinct dc-like second-order nonlinear current when compared to the independent-particle approximation.

Importantly, the qualitative discrepancy in the optical response behavior between the mean-field theories and the independent-particle approximation is not due to excitonic contributions, since the TDH method does not incorporate excitonic effects. This discrepancy emerges solely from the time-dependence of mean fields.

Based on the classification of nonlinear responses outlined in Eqs.~(\ref{eq:opt-rec-dc}--\ref{eq:inj-curr-dc}), a finite dc-like current after laser fields end implies the presence of the injection current. However, it has been extensively discussed that the injection current originates from the breaking of time-reversal symmetry~\cite{PhysRevB.61.5337,10.1063/5.0101513}, such as in the irradiation of circularly polarized light. The residual current depicted in Fig.~\ref{fig:current_t}, resultant from the irradiation with linearly polarized light, is expected to be an artifact, connected to the unphysical divergence of the response function.

A few decades ago, there was considerable discussion regarding the presence of unphysical divergences in second-order nonlinear optical responses in materials, and it has been suggested that these divergences are caused by numerical errors and could be removed by employing a sum-rule suitable for semiconducting systems within the independent particle approximation~\cite{PhysRevB.43.8990}. Another theoretical method involves the evaluation of second-order nonlinear optical responses utilizing an explicit susceptibility expression derived from perturbation theory within the independent particle approximation~\cite{PhysRevB.61.5337}. This perturbative approach successfully avoids the unphysical divergence as it analytically manages the susceptibility singularity.

Even though second-order nonlinear optical phenomena hold considerable importance, studies exploring the divergent behavior of the response function for many-body systems beyond the independent particle approximation have been limited~\cite{PhysRevB.52.14636}. In the following section, we examine the effects of many-body and mean-field effects on the second-order nonlinear optical phenomena based on the perturbation theory.

\section{Perturbation analysis \label{sec:perturbation}}

In the previous section we demonstrated numerically that mean-field approximations, including TDHF and TDH methods, may induce unphysical dc-like current as a result of the second-order nonlinear optical process with linearly polarized light. In this section, we investigate the origin of this unphysical current within perturbation theory. We analyze the current induced by a laser pulse and evaluate dc-like component of the induced current after laser irradiation. This perturbation analysis is performed at the exact many-body Schr\"odinger equation, the independent particle approximation, and the mean-field approximations.

\subsection{Exact Schr\"odinger Equation \label{subsec:exact-tdse}}

We first consider the light-induced current within the exact Schr\"odinger equation for a many-electron system,
\begin{equation}
i\hbar \frac{\partial}{\partial t} \big | \Psi(t) \big \rangle = \hat H(t) \big | \Psi(t) \big \rangle,
\label{eq:many-tdse-org}
\end{equation}
where $\hat H(t)$ denotes the many-electron Hamiltonian
\begin{equation}
\hat H(t)=\sum_i \left [\frac{1}{2m_e}\left( \vecb p_i + e\vecb A(t) \right)^2 +v(\vecb r_i)\right ]+\frac{1}{2}\sum_{ij}w(\vecb r_i - \vecb r_j)
\label{eq:many-ham}
\end{equation}
with a one-body potential $v(\vecb r)$, and an interacting potential $w(\vecb r- \vecb r')$. Here, the time-dependent vector potential $\vecb A(t)$ is included to describe external electric fields as $\vecb E(t)=-\dot{\vecb A}(t)$. For practical analysis with a finite laser pulse, we impose that the vector potential vanishes for $t>t_f$: $\vecb A(t>t_f)=0$.

The time-dependent Hamiltonian for a many-electron system, $\hat H(t)$, can be decomposed into the following three components:
\begin{align}
\hat H(t) = \hat H_0 + \hat V^{(1)}(t) + C(t),
\end{align}
\begin{align}
\hat H_0 = \sum_i \left [ \frac{p^2_i}{2m_e}+v(\vecb r_i)\right ] +\frac{1}{2}\sum_{ij}w(\vecb r_i - \vecb r_j), 
\end{align}
\begin{align}
\hat V(t) &= \sum_i \frac{e\vecb p_i \cdot \vecb A(t)}{m_e} = \frac{e}{m_e}\vecb P\cdot \vecb A(t),
\end{align}
\begin{align}
C(t) = \frac{e^2N_e}{2m_e}\left |\vecb A(t) \right |^2,
\end{align}
where $N_e$ is the number of electrons in the system, and $\vecb P$ is the total momentum of the system given by $\vecb P=\sum_j\vecb p_j$.

For later convenience, we introduce a wavefunction $|\tilde \Psi(t)\rangle$ through a unitary transformation as $|\tilde \Psi(t)\rangle=e^{i\int^t dt' C(t')/\hbar}|\Psi(t)\rangle$. Utilizing this, the many-electron Schr\"odinger equation, Eq.~(\ref{eq:many-tdse-org}), can be rewritten as
\begin{equation}
i\hbar \frac{\partial}{\partial t} \big | \tilde \Psi(t) \big \rangle = \left [\hat H_0 + \hat V(t) \right ]\big | \tilde \Psi(t) \big \rangle.
\label{eq:many-tdse-mod}
\end{equation}

To investigate second-order nonlinear responses, we begin by introducing a perturbative expansion of the wavefunction $|\tilde \Psi(t)\rangle$ up to the second order of perturbation~\cite{RevModPhys.44.602}:
\begin{widetext}
\begin{align}
|\tilde \Psi(t)\rangle = \exp
\left [ -\frac{i}{\hbar}E_0 t-\frac{i}{\hbar}\int^t dt' E^{(1)}(t')-\frac{i}{\hbar}\int^tdt'E^{(2)}(t')
\right ]
\left [
|\Phi_0\rangle + |\delta \Psi^{(1)}(t)\rangle + |\delta \Psi^{(2)}(t)\rangle
\right ],
\label{eq:many-body-wf-ansatz}
\end{align}
\end{widetext}
where $|\Phi_0\rangle$ refers to the ground state wavefunction of the unperturbed Hamiltonian, $\hat H_0$, and $E_0$ represents its associated ground state energy, fulfilling $\hat H_0|\Phi_0\rangle=E_0|\Phi_0\rangle$. The first- and second-order wavefunctions are denoted as $|\delta \Psi^{(1)}(t)\rangle$ and $|\delta \Psi^{(2)}(t)\rangle$, respectively. The corresponding first- and second-order dynamical phase factors are determined by the first-order energy shift, $E^{(1)}(t)$, and the second-order energy shifts, $E^{(2)}(t)$, respectively.

By substituting Eq.~(\ref{eq:many-body-wf-ansatz}) into the modified time-dependent Schr\"odinger equation, Eq.~(\ref{eq:many-tdse-mod}), we derive the following relation for each order:
\begin{widetext}
\begin{align}
i\hbar \frac{\partial}{\partial t}|\delta \Psi^{(1)}(t)\rangle + E^{(1)}(t)|\Phi_0\rangle
&= \left (\hat H_0 -E_0 \right )|\delta \Psi^{(1)}(t)\rangle + \hat V(t)|\Phi_0\rangle, \label{eq:many-body-eom-1st-order-wf} \\
i\hbar \frac{\partial}{\partial t}|\delta \Psi^{(2)}(t)\rangle + E^{(1)}(t)|\delta \Psi^{(1)}(t)\rangle +E^{(2)}(t)|\Phi_0\rangle &= \left (\hat H_0-E_0 \right )|\delta \Psi^{(2)}(t)\rangle + \hat V(t)|\delta \Psi^{(1)}(t)\rangle. \label{eq:many-body-eom-2nd-order-wf}
\end{align}
\end{widetext}

To proceed with the analysis, we introduce the eigenstates of the unperturbed Hamiltonian, $\hat H_0$, as follows:
\begin{align}
\hat H_0 |\Phi_a\rangle = E_a |\Phi_a\rangle.
\label{eq:many-body-tdse-eigenstates}
\end{align}

If there is a set of degenerate eigenstates with respect to Eq.~(\ref{eq:many-body-tdse-eigenstates}), we choose to define the eigenstates such that at least one of the interested Cartesian components of $\vecb P$ is diagonalized within the subspace spanned by these degenerate eigenstates.

By utilizing these eigenstates, the perturbative wavefunctions, $|\delta \Psi^{(1)}(t)\rangle$ and $|\delta \Psi^{(2)}(t)\rangle$, can be expanded as
\begin{align}
|\delta \Psi^{(1)}(t)\rangle &= \sum_{a\neq 0}C^{(1)}_a(t)e^{-i\Omega_a t}|\Phi_a\rangle, \label{eq:many-body-tdse-1st-order-expansion} \\
|\delta \Psi^{(2)}(t)\rangle &= \sum_{a\neq 0}C^{(2)}_a(t)e^{-i\Omega_a t}|\Phi_a\rangle, \label{eq:many-body-tdse-2nd-order-expansion}
\end{align}
where $\Omega_a$ is defined as $\Omega_a=(E_a-E_0)/\hbar$. Here, $C^{(1)}_a(t)$ and $C^{(2)}_a(t)$ represent the expansion coefficients for the first- and second-order wavefunctions, respectively. The expansion excludes the unperturbed ground state $|\Phi_0\rangle$, given that the energy shifts, $E^{(1)}(t)$ and $E^{(2)}(t)$, are defined as
\begin{align}
E^{(1)}(t)&=\langle \Phi_0 | \hat V(t)| \phi_0 \rangle, \\
E^{(2)}(t)&=\langle \Phi_0 | \hat V(t)| \delta \Psi^{(1)}(t) \rangle.
\end{align}

We proceed by substituting Eq.~(\ref{eq:many-body-tdse-1st-order-expansion}) and Eq.~(\ref{eq:many-body-tdse-2nd-order-expansion}) into Eq.~(\ref{eq:many-body-eom-1st-order-wf}) and Eq.~(\ref{eq:many-body-eom-2nd-order-wf}), respectively. Consequently, the derived equations of motion for the expansion coefficients are as follows:
\begin{align}
i\hbar \frac{d}{dt}C^{(1)}_a(t)&=e^{i\Omega_a t}\langle\Phi_a| \hat V(t)| \Phi_0\rangle, \label{eq:many-body-tdse-1st-order-coeff-eom} \\
i\hbar \frac{d}{dt}C^{(2)}_a(t)&=e^{i\Omega_a t}\langle\Phi_a| 
\left (\hat V(t)-E^{(1)}(t) \right )|\delta \Psi^{(1)}(t)\rangle.
\end{align}

From these derived expressions, it is evident that the coefficients $C^{(1)}_a(t)$ and $C^{(2)}_a(t)$ become time-invariant once the perturbation ends ($\hat V(t)=\vecb A(t)=0$ for $t>t_f$).

By using the above perturbative expansion, the second order nonlinear current can be expressed as
\begin{align}
\vecb J^{(2)}(t)&=-\frac{e}{m_e} \langle \delta \Psi^{(1)}(t)| \vecb P |\delta \Psi^{(1)}(t)\rangle \nonumber \\
&-\frac{e}{m_e}\langle\Phi_0| \vecb P |\delta \Psi^{(2)}(t)\rangle +\text{c.c.} \nonumber \\
&=-\frac{e}{m_e} \sum_{n,m}C^{(1),*}_n(t)C^{(1)}_m(t)e^{-\frac{i}{\hbar}(E_m-E_n)t}\langle\Phi_n|\vecb P|\Phi_m\rangle \nonumber \\
&-\frac{e}{m_e}\sum_n C^{(2)}_n(t)e^{-\frac{i}{\hbar}E_nt}\langle\Phi_0|\vecb P|\Phi_n\rangle + \text{c.c.}
\end{align}

Given that $C^{(1)}_a(t)$ and $C^{(2)}_a(t)$ become constant after the perturbation ends ($\hat V(t>t_f)=\vecb A(t>t_f)=0$), we can evaluate the dc-component of the current $\vecb J^{(2)}(t)$ after the fields end as
\begin{align}
\vecb J^{(2)}_{\text{dc}}& =\lim_{T\rightarrow \infty}\frac{1}{T}\int^{t_f+T}_{t_f}dt \vecb J^{(2)}(t) \nonumber \\
&= -\frac{e}{m_e}\sum_a |C^{(1)}_a(t_f)|^2\langle\Phi_a|\vecb P|\Phi_a\rangle.
\label{eq:many-body-tdse-dc-current-definition}
\end{align}

In order to analyze the dc component of the second-order current, we turn our attention to the first-order coefficient, $C^{(1)}_a(t)$, at time $t=t_f$. This is calculated by integrating Eq.~(\ref{eq:many-body-tdse-1st-order-coeff-eom}) as follows:
\begin{align}
C^{(1)}_a(t_f)&=\frac{1}{i\hbar}\int^{t_f}_{-\infty}dt' e^{i\Omega_a t'}
\big \langle \Phi_a\big | \hat V(t') \big | \Phi_0\big \rangle \nonumber \\
&=
\frac{e}{m_e}\frac{1}{i\hbar}\int^{\infty}_{-\infty}dt' e^{i\Omega_a t'} \vecb A(t')\cdot
\big \langle \Phi_a\big | \vecb P \big | \Phi_0\big \rangle \nonumber \\
&=\frac{e}{m_e} \frac{1}{i\hbar}\tilde{\vecb A}\left (\Omega_a \right )\cdot 
\big \langle \Phi_a\big | \vecb P \big | \Phi_0\big \rangle,
\end{align}
where $\tilde{\vecb A}(\omega)$ denotes the Fourier transform of $\vecb A(t)$. Here, we have exploited the property that $\vecb A(t)=0$ for $t>t_f$. By employing this explicit expression of $C^{(1)}_a(t_f)$, the dc component of the current $\vecb J^{(2)}_{\text{dc}}$ in Eq.~(\ref{eq:many-body-tdse-dc-current-definition}) can be evaluated as
\begin{align}
\vecb J^{(2)}_{\text{dc}}=-\frac{e}{m_e} \sum_a \left |
\frac{e}{m_e}\frac{1}{i\hbar}\tilde{\vecb A}\left (\Omega_a \right )\cdot 
\big \langle \Phi_a\big | \vecb P \big | \Phi_0\big \rangle
\right |^2\langle\Phi_a|\vecb P|\Phi_a\rangle.
\label{eq:many-body-tdse-dc-current-with-explicit-expression}
\end{align}

To further analyze the dc-component of the second-order current, we next explore a characteristic of the eigenstates $|\Phi_a\rangle$. The real-space representation of a many-body state $|\Phi_a \rangle$ is defined as $\Phi_a(\vecb r_1, \cdots, \vecb r_{N_e})= \langle \bar{\vecb r} |\Phi_a\rangle$. As a result, the static Schr\"odinger equation with the unperturbed Hamiltonian can be reexpressed as follows:
\begin{widetext}
\begin{align}
\left [\sum_j \left \{ -\frac{\hbar^2}{2m_e} \nabla^2 +v(\vecb r_j) \right \}
+\frac{1}{2}\sum_{i \neq j}w(\vecb r_i - \vecb r_j)
\right ]\Phi_a(\vecb r_1, \cdots, \vecb r_{N_e})
=E_a\Phi_a(\vecb r_1, \cdots, \vecb r_{N_e}).
\label{eq:many-body-se-real-space}
\end{align}
\end{widetext}

It can be readily verified that the complex conjugate of an eigenstate, $\Phi^*_a(\vecb r_1, \cdots, \vecb r_{N_e})$, satisfies Eq.~(\ref{eq:many-body-se-real-space}), indicating that $\Phi^*_a(\vecb r_1, \cdots, \vecb r_{N_e})$ is also an eigenstate of the Hamiltonian. We introduce a \textit{ket} vector, $|\Phi^{(-)}_a\rangle$, to represent the abstract state vector associated with $\Phi^*_a(\vecb r_1, \cdots, \vecb r_{N_e})=\langle \bar{\vecb r}|\Phi^{(-)}_a\rangle$. It is worth noting that $|\Phi^{(-)}_a\rangle$ is the time-reversed state of $|\Phi_a\rangle$, and it may or may not be identical to the original state, $|\Phi_a\rangle$.

Reflecting the time-reversal nature, the expectation values of the total momentum calculated with a time-reversed state, $|\Phi^{(-)}_a\rangle$, and the original state, $|\Phi_a\rangle$, have the opposite signs as
\begin{align}
\langle \Phi_{a}|\vecb P|\Phi_{a}\rangle=-\langle \Phi^{(-)}_{a}|\vecb P|\Phi^{(-)}_{a}\rangle.
\label{eq:many-body-time-reversal-matrix-elements}
\end{align}

Moreover, by assuming that the ground state has the time-reversal symmetry as $|\Phi^{(-)}_{0}\rangle=|\Phi_{0}\rangle$, matrix elements of the total momentum operator hold the following relation:
\begin{align}
\langle \Phi_{a}|\vecb P|\Phi_{0}\rangle=-\langle \Phi^{(-)}_{a}|\vecb P|\Phi_{0}\rangle^*.
\label{eq:many-body-time-reversal-matrix-elements-gs}
\end{align}

By employing Eq.~(\ref{eq:many-body-time-reversal-matrix-elements}) and Eq.~(\ref{eq:many-body-time-reversal-matrix-elements-gs}), the dc-component of the second-order nonlinear current in Eq.~(\ref{eq:many-body-tdse-dc-current-with-explicit-expression}) can be further evaluated with the time-reversed states as
\begin{widetext}
\begin{align}
\vecb J^{(2)}_{\text{dc}}&=-\frac{1}{2}\frac{e}{m_e} \left [
\sum_a \left |
\frac{e}{m_e}\frac{1}{i\hbar}\tilde{\vecb A}\left (\Omega_a \right )\cdot 
\big \langle \Phi_a\big | \vecb P \big | \Phi_0\big \rangle
\right |^2\langle\Phi_a|\vecb P|\Phi_a\rangle
+\sum_a
\left |
\frac{e}{m_e}\frac{1}{i\hbar}\tilde{\vecb A}\left (\Omega_a \right )\cdot 
\big \langle \Phi^{(-)}_a\big | \vecb P \big | \Phi_0\big \rangle
\right |^2\langle\Phi^{(-)}_a|\vecb P|\Phi^{(-)}_a\rangle
\right ] \nonumber \\
&= -\frac{e}{2m_e}\sum_a \left [
\left |\frac{e}{m_e}
\frac{1}{i\hbar}\tilde{\vecb A}\left (\Omega_a \right )\cdot 
\big \langle \Phi_a\big | \vecb P \big | \Phi_0\big \rangle
\right |^2\langle\Phi_a|\vecb P|\Phi_a\rangle
+\left |\frac{e}{m_e}
\frac{1}{i\hbar}\tilde{\vecb A}\left (\Omega_a \right )\cdot 
\big \langle \Phi^{(-)}_a\big | \vecb P \big | \Phi_0\big \rangle
\right |^2\langle\Phi^{(-)}_a|\vecb P|\Phi^{(-)}_a\rangle 
\right ] \nonumber \\
&=-\frac{e^3}{2m_e^3}\sum_a \langle\Phi_a|\vecb P|\Phi_a\rangle \left [
\left |
\frac{1}{i\hbar}\tilde{\vecb A}\left (\Omega_a \right )\cdot 
\big \langle \Phi_a\big | \vecb P \big | \Phi_0\big \rangle
\right |^2
-
\left |
\frac{1}{i\hbar}\tilde{\vecb A}\left (\Omega_a \right )\cdot 
\big \langle \Phi_a\big | \vecb P \big | \Phi_0\big \rangle^*
\right |^2
\right ] \nonumber \\
&=-\frac{e^3}{2m_e^3 \hbar^2}\sum_a \langle\Phi_a|\vecb P|\Phi_a\rangle \left [
\left |
\tilde{\vecb A}\left (\Omega_a \right )\cdot 
\big \langle \Phi_a\big | \vecb P \big | \Phi_0\big \rangle
\right |^2
-
\left |
\tilde{\vecb A}^*\left (\Omega_a \right )\cdot 
\big \langle \Phi_a\big | \vecb P \big | \Phi_0\big \rangle
\right |^2
\right ].
\label{eq:many-body-dc-current-final} 
\end{align}
\end{widetext}

The final expression for the dc-component of the current, Eq.~(\ref{eq:many-body-dc-current-final}), suggests that residual dc-component cannot be induced solely by linearly polarized light as a second-order nonlinear effect. This can be straightforwardly confirmed by evaluating Eq.~(\ref{eq:many-body-dc-current-final}) by employing a linearly polarized vector potential in the frequency domain, $\tilde{\vecb A}(\omega)=\tilde A(\omega)\vecb e_p$, with a pure-real unit vector along the polarization direction, resulting in $\vecb J^{(2)}_{\text{dc}}=0$.

By contrast, if we evaluate Eq.~(\ref{eq:many-body-dc-current-final}) with elliptically polarized light represented as $\tilde {\vecb A}(\omega)=\vecb e_x \tilde A_x(\omega)+\vecb e_y \tilde A_y(\omega)$, the dc-component is given by
\begin{align}
&\vecb J^{(2)}_{\text{dc}}=-\frac{e^3}{m_e^3 \hbar^2}\Re  \Big [\sum_a 
\langle \Phi_{a}\big | \vecb P \big | \Phi_{a}\big \rangle
\big \langle \Phi_{a}\big | P_x \big | \Phi_0\big \rangle \nonumber \\
&\times \big \langle \Phi_{a}\big | P_y \big | \Phi_0\big \rangle^*
\left \{
\tilde A_x (\Omega_{a} )
\tilde A_y^*(\Omega_{a} )
-\tilde A_x^*(\Omega_{a} )
\tilde A_y(\Omega_{a} )
\right \}
\Big ].
\end{align}

This result indicates that the dc-component of the current may be induced only if the light field is elliptically polarized ($\tilde A_x (\Omega_{a} )\tilde A_y^*(\Omega_{a})-\tilde A_x^*(\Omega_{a})\tilde A_y(\Omega_{a})\neq 0$), or namely breaking the time-reversal symmetry. This current represents the injection current induced by a time-reversal symmetry broken field in a system lacking inversion symmetry.

Working with the many-body Schr\"odinger equation, we have confirmed that the second-order dc current cannot exist after a laser pulse ends if the laser field is linearly polarized in a time-reversal symmetry system. In other words, the injection current cannot be induced by a linearly polarized light, but it can be induced only by an elliptically polarized light. This conclusion is consistent with previous works~\cite{PhysRevB.61.5337,10.1063/5.0101513}.

\subsection{Mean-field theory \label{subsec:mean-field-tdse}}

To extend the perturbation analysis with the exact many-body Schr\"odinger equation, here we analyze the dc component of the second-order nonlinear current by employing mean-field theories. As a practical mean-field approximation to solid-state systems, we assume that a many-electron system is described by a single Slater determinant, and each electronic orbital is given by a Bloch state, $\psi_{b\vecb k}(\vecb r,t)=e^{i(\vecb k + \vecb A(t)/\hbar)\cdot \vecb r}u_{b\vecb k}(\vecb r,t)$, where $u_{b\vecb k}(\vecb r,t)=u_{b\vecb k}(\vecb r+\vecb a_j,t)$ stands for the periodic part of the Bloch function and $\vecb a_j$ is any lattice vectors. Furthermore, we impose that electronic orbitals, $\psi_{b\vecb k}(\vecb r,t)$, obey the following mean-field equation of motion:
\begin{align}
i\hbar \frac{\partial}{\partial t}\psi_{b\vecb k}(\vecb r,t)&=\hat h(t)\psi_{b\vecb k}(\vecb r,t) \nonumber \\
&=\left [ \frac{\left \{\vecb p +e\vecb A(t) \right \}^2}{2m}+\hat v(t) \right ] \psi_{b\vecb k}(\vecb r,t),
\label{eq:mf-eom}
\end{align}
where $\hat v(t)$ denotes a mean-field potential operator. We assume that the Hamiltonian $\hat h(t)$ has spatial periodicity with the lattice vectors, $\vecb a_j$.

Both the TDHF and TDH methods utilized in the numerical analysis in Section~\ref{sec:results} are described by the same form as Eq.~(\ref{eq:mf-eom}). Furthermore, it is worth noting that the time-dependent Kohn--Sham equation in the time-dependent density functional theory is also described by the same form.

In order to proceed with the perturbation analysis, we introduce the eigenstates of the unperturbed Hamiltonian $\hat{h}_0$ as follows:
\begin{align}
\hat{h}_0 \phi_{b\vecb{k}}(\vecb{r}) &= \epsilon_{b\vecb{k}} \phi_{b\vecb{k}}(\vecb{r}), \\
\hat{h}_0 &= \frac{\vecb{p}^2}{2m} + \hat{v}_0,
\end{align}
where $\hat{h}_0$ is the unperturbed Hamiltonian, $\phi_{b\vecb{k}}(\vecb{r})$ represents the eigenstates of $\hat{h}_0$, and $\hat{v}_0$ is the non-perturbed potential. According to the discussion for the perturbation analysis with the exact many-body Schr\"odinger equation, if there is a set of degenerate eigenstates, we choose to define the eigenstates such that at least one of the Cartesian components of $\vecb p$ is diagonalized within the subspace spanned by these degenerate eigenstates.

Next, we expand the time-dependent Hamiltonian $\hat{h}(t)$ up to the second order in terms of the external field $\vecb{A}(t)$ as follows:
\begin{align}
\hat{h}(t) &= \hat{h}_0 + \hat{h}^{(1)}(t) + \hat{h}^{(2)}(t), \\
\hat{h}^{(1)}(t) &= \frac{e\vecb{p} \cdot \vecb{A}(t)}{m} + \delta \hat{v}^{(1)}(t), \\
\hat{h}^{(2)}(t) &= \frac{e^2\vecb{A}^2(t)}{2m} + \delta \hat{v}^{(2)}(t),
\end{align}
where $\delta \hat{v}^{(1)}(t)$ and $\delta \hat{v}^{(2)}(t)$ represent the first- and second-order contributions from the mean-field potential $\hat{v}(t)$, respectively.

Assuming the time-dependent wavefunctions $\psi_{b\vecb k}(\vecb r, t)$ to be initially prepared as the eigenstates of $\hat{h}_0$ i.e. $\psi_{b\vecb k}(\vecb r, t=-\infty) = \phi_{b\vecb k}(\vecb r)$, we can expand $\psi_{b\vecb k}(\vecb r, t)$ up to the second order of the external field as follows:
\begin{align}
\psi_{b\vecb k}(\vecb r, t) &= \exp\left[-\frac{i}{\hbar}\int^t dt' \epsilon_{b\vecb k} + \delta \epsilon^{(1)}_{b\vecb k}(t') + \delta \epsilon^{(2)}_{b\vecb k}(t')\right] \nonumber \\
& \times \left[\phi_{b\vecb k}(\vecb r) + \delta \psi^{(1)}_{b\vecb k}(\vecb r, t) + \delta \psi^{(2)}_{b\vecb k}(\vecb r, t)\right].
\label{eq:mf-wf-perturbed-expansion}
\end{align}

To ensure the orthogonality relations, $\langle \phi_{b\vecb k}|\delta \psi^{(1)}_{b\vecb k}(t) \rangle = \langle \phi_{b\vecb k}|\delta \psi^{(2)}_{b\vecb k}(t) \rangle = 0$, we choose the energy shifts as follows:
\begin{align}
\delta \epsilon^{(1)}_{b\vecb k}(t) &= \langle \phi_{b\vecb k}| \hat{h}^{(1)}(t)| \phi_{b\vecb k}\rangle, \\
\delta \epsilon^{(2)}_{b\vecb k}(t) &= \langle \phi_{b\vecb k}| \hat{h}^{(2)}(t)| \phi_{b\vecb k}\rangle + \langle \phi_{b\vecb k}| \hat{h}^{(1)}(t)| \delta \psi^{(1)}_{b\vecb k}\rangle.
\end{align}

Furthermore, we expand $\delta \psi^{(1)}_{b\vecb{k}}(\vecb{r},t)$ and $\delta \psi^{(2)}_{b\vecb{k}}(\vecb{r},t)$ in terms of the eigenstates $\phi_{b\vecb{k}}(\vecb{r})$ as follows:
\begin{align}
\delta \psi^{(1)}_{b\vecb{k}}(\vecb{r},t) &= \sum_{a \neq b} c^{(1)}_{a,b\vecb{k}}(t) \phi_{a\vecb{k}}(\vecb{r}) e^{-i\omega_{ab\vecb{k}}t}, \label{eq:mf-1st-wf-basis-exp} \\
\delta \psi^{(2)}_{b\vecb{k}}(\vecb{r},t) &= \sum_{a \neq b} c^{(2)}_{a,b\vecb{k}}(t) \phi_{a\vecb{k}}(\vecb{r}) e^{-i\omega_{ab\vecb{k}}t}, \label{eq:mf-2nd-wf-basis-exp}
\end{align}
where $c^{(1)}_{a,b\vecb{k}}(t)$ and $c^{(2)}_{a,b\vecb{k}}(t)$ represent the first- and second-order expansion coefficients, respectively. The frequency $\omega_{ab\vecb{k}}$ is defined as $\omega_{ab\vecb{k}} = (\epsilon_{a\vecb{k}} - \epsilon_{b\vecb{k}}) / \hbar$.

By substituting Eqs.~(\ref{eq:mf-wf-perturbed-expansion}), (\ref{eq:mf-1st-wf-basis-exp}), and (\ref{eq:mf-2nd-wf-basis-exp}) into Eq. (\ref{eq:mf-eom}), we can derive the equations of motion for the coefficients $(a\neq b)$ as follows:
\begin{align}
i\hbar \frac{d}{dt}c^{(1)}_{a,b\vecb{k}}(t) &= e^{i\omega_{ab\vecb{k}}t} \int d\vecb{r} \phi^*_{a\vecb{k}}(\vecb{r}) \hat{h}^{(1)}(t) \phi_{b\vecb{k}}(\vecb{r}), \label{eq:mf-1st-coeff-eom} \\
i\hbar \frac{d}{dt}c^{(2)}_{a,b\vecb{k}}(t) &= e^{i\omega_{ab\vecb{k}}t} \int d\vecb{r} \phi^*_{a\vecb{k}}(\vecb{r}) \hat{h}^{(2)}(t) \phi_{b\vecb{k}}(\vecb{r}) \nonumber \\
&+ e^{i\omega_{ab\vecb{k}}t} \int d\vecb{r} \phi^*_{a\vecb{k}}(\vecb{r}) \hat{h}^{(1)}(t) \delta\psi^{(1)}_{b\vecb{k}}(\vecb{r},t) \nonumber \\
&- \delta\epsilon^{(1)}_{b\vecb{k}}(t) c^{(1)}_{a,b\vecb{k}}(t). \label{eq:mf-2nd-coeff-eom}
\end{align}

It is worth noting that the coefficients, $c^{(1)}_{a,b\vecb{k}}(t)$ and $c^{(2)}_{a,b\vecb{k}}(t)$, may change only in the presence of $\hat{h}^{(1)}(t)$ or $\hat{h}^{(2)}(t)$.

\subsubsection{Independent particle approximation} \label{subsubsec:linear-ip}

To highlight contributions from the time-dependent mean-field to the dc-component of nonlinear current, we first revisit the results of the independent particle approximation, obtained by setting $\delta \hat v^{(1)}(t) = \delta \hat v^{(2)}(t) = 0$. Under these constraints, the coefficients $c^{(1)}_{a,b\vecb{k}}(t)$ and $c^{(2)}_{a,b\vecb{k}}(t)$ remain time-invariant after the laser irradiation since both $\hat{h}^{(1)}(t)$ and $\hat{h}^{(2)}(t)$ become zero. For practical analysis, we assume $\vecb{A}(t) = 0$ for $t > t_f$. Consequently, the first-order coefficient can be expressed as
\begin{align}
c^{(1)}_{a,b\vecb{k}}(t \ge t_f) = \frac{1}{i\hbar}\frac{e}{m} \tilde{\vecb{A}}(\omega_{ab\vecb{k}}) \cdot \int d\vecb{r} \phi^{*}_{a\vecb{k}} \vecb{p} \phi_{b\vecb{k}}(\vecb{r}).
\label{eq:mf-ip-coeff-1st}
\end{align}

Next, we calculate the second-order nonlinear current associated with the orbital $\psi_{b\vecb{k}}(\vecb{r}, t)$ as follows:
\begin{align}
\vecb{J}^{(2)}_{b\vecb{k}}(t) &=-\frac{e}{m} \int d\vecb{r} \delta \psi^{(1),*}_{b\vecb{k}}(\vecb{r}, t) \vecb{p} \delta \psi^{(1)}_{b\vecb{k}}(\vecb{r}, t) \nonumber \\
&-\frac{e}{m} \int d\vecb{r} \phi^{*}_{b\vecb{k}} \vecb{p} \delta \psi^{(2)}_{b\vecb{k}}(\vecb{r}, t) + \text{c.c.}
\end{align}

As discussed in Sec.~\ref{subsec:exact-tdse}, we can determine the dc-component of the second-order nonlinear current after the laser field ends using the following expression:
\begin{widetext}
\begin{align}
\vecb J^{(2)}_{b\vecb k,\dc} &= \lim_{T\rightarrow \infty} \frac{1}{T} \int^{t_f+T}_{t_f}dt \vecb J^{(2)}_{b\vecb k}(t) = -\frac{e}{m} \sum_a \left |c^{(1)}_{a,b\vecb k}(t>t_f) \right |^2 \int d\vecb r \phi^{*}_{a\vecb k}\vecb p \phi_{a\vecb k}(\vecb r) \nonumber \\
&= -\frac{e}{m} \sum_a \int d\vecb r \phi^{*}_{a\vecb k}\vecb p \phi_{a\vecb k}(\vecb r) \left |\frac{1}{i\hbar}\frac{e}{m} \tilde {\vecb A}\left (\omega_{ab\vecb k} \right )\cdot \int d\vecb r \phi^{*}_{a\vecb k}\vecb p \phi_{b\vecb k}(\vecb r) \right |^2.
\label{eq:mf-dc-current-bk}
\end{align}

Assuming that the unperturbed mean-field Hamiltonian $\hat{h}_0$ has time-reversal symmetry, the Bloch states at $\vecb{k}$ and $-\vecb{k}$ exhibit the following time-reversal relations: $\phi_{b,-\vecb{k}}(\vecb{r})=\phi_{b\vecb{k}}^*(\vecb{r})$ and $\epsilon_{b,-\vecb{k}}=\epsilon_{b\vecb{k}}$. By utilizing these relations, the sum of the dc-component of the current at $\vecb{k}$ and $-\vecb{k}$ can be evaluated as
\begin{align}
\vecb{J}^{(2)}_{b\vecb{k},\mathrm{dc}}+\vecb{J}^{(2)}_{b,-\vecb{k},\mathrm{dc}}&=
-\frac{e}{m}\sum_a \Bigg[\bigg |c^{(1)}_{a,b\vecb{k}}(t>t_f) \bigg |^2
\int d\vecb{r} \phi^{*}_{a\vecb{k}}\vecb{p} \phi_{a\vecb{k}}(\vecb{r})
 +\bigg |c^{(1)}_{a,b,-\vecb{k}}(t>t_f) \bigg |^2
\int d\vecb{r} \phi^{*}_{a,-\vecb{k}}\vecb{p} \phi_{a,-\vecb{k}}(\vecb{r})
\Bigg]\nonumber \\
&=-\frac{e}{m}\sum_a\int d\vecb{r} \phi^{*}_{a\vecb{k}}\vecb{p} \phi_{a\vecb{k}}(\vecb{r}) 
\bigg[\bigg |c^{(1)}_{a,b\vecb{k}}(t>t_f) \bigg |^2- \bigg |c^{(1)}_{a,b,-\vecb{k}}(t>t_f) \bigg |^2
\bigg]\nonumber \\
&=-\frac{e^3}{m^3\hbar ^2}\sum_a \int d\vecb{r} \phi^{*}_{a\vecb{k}}\vecb{p} \phi_{a\vecb{k}}(\vecb{r}) 
\times \bigg[
\bigg|\tilde{\vecb{A}}\left(\omega_{ab\vecb{k}}\right) \cdot
\int d\vecb{r} \phi^{*}_{a\vecb{k}}\vecb{p} \phi_{b\vecb{k}}(\vecb{r}) \bigg|^2
 -\bigg|\tilde{\vecb{A}}^*\left(\omega_{ab\vecb{k}}\right) \cdot
\int d\vecb{r} \phi^{*}_{a\vecb{k}}\vecb{p} \phi_{b\vecb{k}}(\vecb{r}) \bigg|^2
\bigg].
\label{eq:mf-dc-current-sum-pk-mp}
\end{align}

Finally, the dc-component of the total current can be evaluated as
\begin{align}
\vecb{J}^{(2)}_{\mathrm{dc}}&=\sum_{b=\mathrm{occ}}\frac{1}{\Omega_{\mathrm{BZ}}} \int_{\Omega_{\mathrm{BZ}}} d\vecb{k}\, \vecb{J}^{(2)}_{b\vecb{k},\mathrm{dc}} \nonumber \\
&=\frac{1}{2}\sum_{b=\mathrm{occ}}\frac{1}{\Omega_{\mathrm{BZ}}} \int_{\Omega_{\mathrm{BZ}}} d\vecb{k} \left(\vecb{J}^{(2)}_{b\vecb{k},\mathrm{dc}}+\vecb{J}^{(2)}_{b,-\vecb{k},\mathrm{dc}} \right) \nonumber \\
&=-\frac{e^3}{2m^3\hbar^2}\sum_{b=\mathrm{occ}}\frac{1}{\Omega_{\mathrm{BZ}}} \int_{\Omega_{\mathrm{BZ}}} d\vecb{k} 
\sum_a \int d\vecb{r} \phi^{*}_{a\vecb{k}}\vecb{p} \phi_{a\vecb{k}}(\vecb{r}) 
 \bigg[
\bigg|\tilde{\vecb{A}}\left(\omega_{ab\vecb{k}}\right) \cdot
\int d\vecb{r} \phi^{*}_{a\vecb{k}}\vecb{p} \phi_{b\vecb{k}}(\vecb{r}) \bigg|^2
-\bigg|\tilde{\vecb{A}}^*\left(\omega_{ab\vecb{k}}\right) \cdot
\int d\vecb{r} \phi^{*}_{a\vecb{k}}\vecb{p} \phi_{b\vecb{k}}(\vecb{r}) \bigg|^2
\bigg],
\label{eq:mf-dc-current-final}
\end{align}
\end{widetext}
where the sum of the index $b$ is taken only for the occupied orbitals ($\mathrm{occ}$), and $\Omega_{\mathrm{BZ}}$ is the volume of the Brillouin zone.

From the final expression in Eq.~(\ref{eq:mf-dc-current-final}), it is evident that the dc-component of the current $\vecb J^{(2)}_{\dc}$ vanishes when linearly polarized light is considered. For instance, assuming a vector potential of the form $\tilde{\vecb A}(\omega)=\tilde A(\omega)\vecb e_p$, $\vecb J^{(2)}_{\dc}$ vanishes due to the integrand of the last line of Eq.~(\ref{eq:mf-dc-current-final}) becoming zero. Instead, under elliptically polarized light, the dc-component of the current may remain finite. If we consider the vector potential to be of the form $\tilde{\vecb A}(\omega)=\tilde{A_x}(\omega)\vecb e_x+\tilde{A_y}(\omega)\vecb e_y$, the dc-component of the current in Eq.~(\ref{eq:mf-dc-current-final}) is given by
\begin{align}
\vecb J^{(2)}_{\dc}&=-\frac{e^3}{2m^3\hbar^2}\sum_{b=occ}\frac{1}{\Omega_{\mathrm{BZ}}} \int_{\Omega_{\mathrm{BZ}}} d\vecb k 
\sum_a \int d\vecb r \phi^{*}_{a\vecb k}\vecb p \phi_{a\vecb k}(\vecb r) \nonumber \\
&\times \left (\int d\vecb r \phi^{*}_{a\vecb k}p_x \phi_{b\vecb k}(\vecb r) \right )
\left (\int d\vecb r \phi^{*}_{a\vecb k}p_y \phi_{b\vecb k}(\vecb r) \right )^* \nonumber \\
&\times \left [\tilde A_x(\omega_{ab\vecb k}) \tilde A^*_y(\omega_{ab\vecb k})-
\tilde A^*_x(\omega_{ab\vecb k})\tilde A_y(\omega_{ab\vecb k})\right ]+c.c.
\label{eq:mf:ip:injection-current}
\end{align}

In contrast to the current under linearly polarized light, Eq.~(\ref{eq:mf:ip:injection-current}) shows that the dc-component of the current may remain finite under elliptically polarized light, indicating the possibility of inducing an injection current. Moreover, the dc-component of the current in Eq.~(\ref{eq:mf:ip:injection-current}) arises from the interference between excited states induced by $\tilde{A}_x(\omega)\vecb e_x$ and $\tilde{A}_y(\omega)\vecb e_y$. This is nothing but the quantum interference among two excitation paths associated with orthogonal components of light fields \cite{PhysRevB.61.5337,10.1063/1.2131191}

\subsubsection{A mean-field approximation with linearly polarized light}

Here, we investigate the impact of a mean-field contribution on the dc-component of the total induced current. To achieve this, we incorporate the contributions from $\hat{v}^{(1)}(t)$ and $\hat{v}^{(2)}(t)$ into the perturbation analysis, employing the independent-particle approximation, as discussed in Sec.~\ref{subsubsec:linear-ip}.

To specifically examine the mean-field contribution only during laser irradiation, we neglect the induced mean-field effect after the laser fields ends. For practical analysis, we impose $\hat{v}^{(1)}(t)=\hat{v}^{(2)}(t)=0$ as well as $\vecb{A}(t)=0$ for $t>t_f$. Under these constraints, we can evaluate the dc-component of the current after the laser fields end as follows:
\begin{align}
\vecb{J}^{(2)}_{\mathrm{dc}} &= \lim_{T \rightarrow \infty}\frac{1}{T}\int^{t_f+T}_{t_f}dt \sum_{b=\mathrm{occ}}\frac{1}{\Omega_{\mathrm{BZ}}}
\int_{\Omega_{\mathrm{BZ}}} d\vecb{k} \vecb{J}^{(2)}_{b\vecb{k}}(t) \nonumber \\
&=-\frac{e}{m} \sum_a\frac{1}{\Omega_{\mathrm{BZ}}}\int_{\Omega_{\mathrm{BZ}}} d\vecb{k}
\int d\vecb{r} \phi^*_{a\vecb{k}}(\vecb{r})\vecb{p} \phi_{a\vecb{k}}(\vecb{r}) \delta n_{a\vecb{k}},
\end{align}
where $\delta n_{a \vecb{k}}$ represents the population imbalance between $\vecb{k}$ and $-\vecb{k}$ and is defined as:
\begin{align}
\delta n_{a \vecb{k}} = \sum_{b=\mathrm{occ}} \left [
 \left |c^{(1)}_{a,b\vecb{k}}(t>t_f) \right |^2
- \left |c^{(1)}_{a,b,-\vecb{k}}(t>t_f) \right |^2
 \right ]. \label{eq:mf-pop-imbalance-def}
\end{align}

By integrating Eq.~(\ref{eq:mf-1st-coeff-eom}) with the mean-field contribution $\delta \hat{v}^{(1)}(t)$, we can evaluate the first-order coefficient as follows:
\begin{align}
c^{(1)}_{a,b\vecb{k}}(t>t_f) &= \frac{1}{i\hbar}\frac{e}{m} \tilde{\vecb{A}}\left(\omega_{ab\vecb{k}}\right) \cdot 
\int d\vecb{r} \, \phi^{*}_{a\vecb{k}} \vecb{p} \, \phi_{b\vecb{k}}(\vecb{r}) \nonumber \\
&+ \frac{1}{i\hbar} \int d\vecb{r} \, \phi^{*}_{a\vecb{k}} \, \tilde{v}^{(1)}\left(\omega_{ab\vecb{k}}\right) \, \phi_{b\vecb{k}}(\vecb{r}), \label{eq:mf-coeff-1st-mf}
\end{align}
where $\tilde{v}^{(1)}(\omega)$ represents the Fourier transform of $\delta \hat{v}^{(1)}(t)$.

By substituting Eq.~(\ref{eq:mf-coeff-1st-mf}) into Eq.~(\ref{eq:mf-pop-imbalance-def}), the population imbalance can be rewritten using the following decomposition:
\begin{align}
\delta n_{a\vecb{k}} = \sum_{b=\text{occ}} \left(\delta n^{A}_{a,b\vecb{k}} + \delta n^{B}_{a,b\vecb{k}} + \delta n^{C}_{a,b\vecb{k}} \right).
\end{align}

Each decomposed component is given as follows:
\begin{widetext}
\begin{align}
\delta n^{A}_{a,b \vecb k}&=\frac{e^2}{m^2\hbar^2} \left |
\tilde {\vecb A}\left (\frac{\epsilon_{a\vecb k}-\epsilon_{b\vecb k}}{\hbar} \right )\cdot
\int d\vecb r \phi^{*}_{a\vecb k}\vecb p \phi_{b\vecb k}(\vecb r)
\right |^2 
-\frac{e^2}{m^2\hbar^2} \left |
\tilde {\vecb A^*}\left (\frac{\epsilon_{a\vecb k}-\epsilon_{b\vecb k}}{\hbar} \right )\cdot
\int d\vecb r \phi^{*}_{a\vecb k}\vecb p \phi_{b\vecb k}(\vecb r)
\right |^2, \\
\delta n^{B}_{a,b \vecb k}&=\frac{2e}{m\hbar^2}\Re \left [
\int d\vecb r \phi^{*}_{a\vecb k}\vecb p \phi_{b\vecb k}(\vecb r)\cdot
\left \{
\tilde {\vecb A}(\omega_{ab\vecb k})
\left (\int d\vecb r \phi^*_{a\vecb k} \tilde{v}^{(1)} (\omega_{ab\vecb k}) \phi_{b\vecb k}(\vecb r)  \right )^*
+\tilde {\vecb A^*}(\omega_{ab\vecb k})
\left (\int d\vecb r \phi_{a\vecb k} \tilde{v}^{(1)} (\omega_{ab\vecb k}) \phi^*_{b\vecb k}(\vecb r)  \right )
\right \} \right ], \\
\delta n^{C}_{a,b \vecb k} &= \frac{1}{\hbar^2}
\left |
\int d\vecb r \phi^*_{a\vecb k} \tilde{v}^{(1)} (\omega_{ab\vecb k}) \phi_{b\vecb k}(\vecb r)
\right |^2
-\frac{1}{\hbar^2}\left |
\int d\vecb r \phi_{a\vecb k} \tilde{v}^{(1)} (\omega_{ab\vecb k}) \phi^*_{b\vecb k}(\vecb r)
\right |^2.
\end{align}
\end{widetext}

In the above expressions, $\delta n^{A}_{a,b \vecb k}$ represents the population imbalance induced by direct excitation from laser fields $\vecb A(t)$. The imbalance $\delta n^{B}_{a,b \vecb k}$ arises from interference between quantum states excited by both the laser fields $\vecb A(t)$ and the induced mean-field $\hat v^{(1)}(t)$. Finally, $\delta n^{C}_{a,b \vecb k}$ is the imbalance generated solely by the mean-field contribution $\hat v^{(1)}(t)$.

Consistently with the exact many-body Schr\"odinger equation and the independent particle approximation, the population imbalance $\delta n^{A}_{a,b \vecb k}$ becomes zero if the external field $\vecb A(t)$ is linearly polarized light, but it may remain finite only if $\vecb A(t)$ is elliptically polarized light. Conversely, $\delta n^{B}_{a,b \vecb k}$ and $\delta n^{C}_{a,b \vecb k}$ may remain finite even under the application of linearly polarized light. The induced mean-field enables additional excitation pathways, leading to quantum interference, which results in the population imbalance and subsequent dc current even after laser irradiation ends. This behavior is consistent with the results of our numerical simulations with the mean-field approximations (TDHF and TDH) shown in Fig.~\ref{fig:current_t}, which also demonstrate dc current remains finite even after the laser fields end. Notably, the independent particle approximation does not show any dc current after the laser fields end in Fig.~\ref{fig:current_t}.

We emphasize that the resultant dc current following the irradiation by linearly polarized light is an artifact of the mean-field approximation. Such a dc current cannot be induced if we consider the exact many-body Schr\"odinger equation (see discussion in Sec.~\ref{subsec:exact-tdse}). This artifact manifests as an unphysical divergence in the nonlinear susceptibility in the low-frequency limit. The unphysical divergence may overcome the intrinsic low-frequency response of the system, inhibiting a proper investigation of the photovoltaic effect, such as shift current, within the scope of mean-field theories unless the artifact is entirely eliminated.

The perturbation analysis suggests that the unphysical dc current originates from excitation paths opened up by the induced mean-field dynamics. Therefore, the artifact is a result of \textit{self-excitation} via the mean-field that are not present in the fully interacting many-body solutoin (see Sec.~\ref{subsec:exact-tdse}). One could potentially improve the mean-field description for the photovoltaic effect by eliminating the unphysical self-excitation effect. This can be achieved by appropriately designing the Hartree-exchange-correlation kernel $f_{\mathrm{Hxc}}(\vecb r, \vecb r', \omega)$ in the time-dependent density functional theory.

\section{First-principles analysis \label{sec:tddft}}

Having established the understanding on the limitation of the mean-field approximation resulting in the unphysical dc current based on the one-dimensional model simulation, we then quantify this limitation for more realistic situations by using the first-principles calculations based on the TDDFT. As an example of realistic systems, we take BaTiO$_3$. For the investigation on the nonlinear current, we first compute the ground state of the tetragonal phase of BaTiO$_3$ with the structural parameters at $300$~K~\cite{Kwei1993} by using the local density approximation \cite{PhysRevB.45.13244}. We note that the polarization direction of the system is the $c$-axis.

For the description of electronic systems, we employ the norm-conserving pseudopotential approximations: For barium atoms, 5s, 5p, and 6s electrons are treated as valence with the multireference pseudopotential scheme \cite{PhysRevB.68.155111,OLIVEIRA2008524}. For titanium, 3s, 3p, 3d and 4s electrons are treated as valence with the multireference pseudopotential scheme \cite{PhysRevB.68.155111,OLIVEIRA2008524}. For oxygen, we employ the Hartwigsen--Goedecker--Hutter (HGH) pseudopotential \cite{PhysRevB.58.3641}. In this work, practical DFT and TDDFT calculations are performed with \textit{Octopus} code \cite{10.1063/1.5142502}. For numerical simulations, the primitive cell of BaTiO$_3$ is discretized into $32^3$ grid points. Similarly, the first Brillouin zone is discretized into $16^3$  $k$-points.

Once the ground state of the tetragonal BaTiO$_3$ is prepared with the above conditions and parameters, we then compute the light-induced electron dynamics by solving the time-dependent Kohn-Sham equation with the adiabatic local density approximation (ALDA), where the exchange-correlation potential is evaluated by the local density approximation with the instantaneous electron density at each time. As an external field, we employ a vector potential polarized along $c$-axis with the form of Eq.~(\ref{eq:laser-pulse-vec}). For practical calculations of BaTiO$_3$, we set $E_0$ to $2.75$~MV/cm, $\omega_0$ to $4$~eV/$\hbar$, and $T_{\mathrm{pulse}}$ to $20$~fs. We repeat the TDDFT calculations with the external fields by using four different CEPs, $\phi_{\mathrm{CEP}}=0, \pi/2, \pi$, and $3\pi/2$. By averaging the obtained current from these calculations, we extract the dc-like current component of the light-induced current with Eq.~(\ref{eq:dc-current-phi-integral}).

\begin{figure}[htb]
\includegraphics[width=0.97\linewidth]{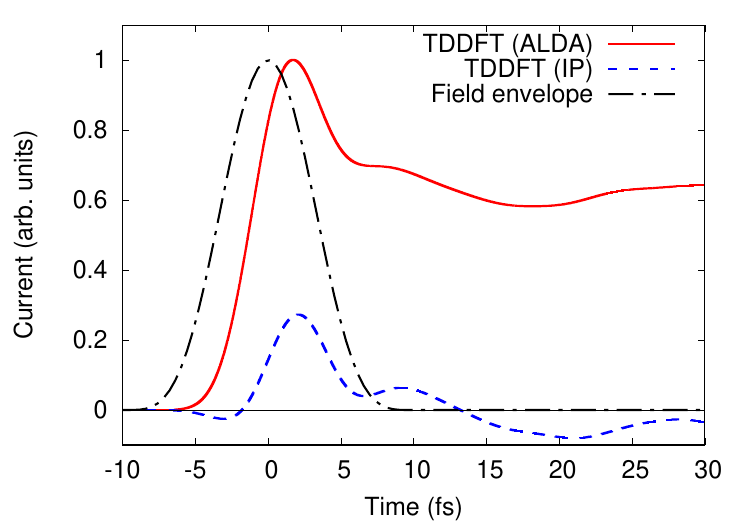}
\caption{\label{fig:shift_current_tddft}
The dc-like current in BaTiO$_3$ induced by a linearly polarized laser pulse. The results are computed by using the TDDFT with the ALDA (red solid line) and with the independent particle approximation (blue dashed line).
}
\end{figure}

Figure~\ref{fig:shift_current_tddft} shows the extracted dc-like current in BaTiO$_3$ as a function of time. As a reference, the envelope function of the applied laser field is also shown as the black dash-dot line. In Fig.~\ref{fig:shift_current_tddft}, the result obtained by using the ALDA is shown as the red solid line, while that obtained by using the independent particle approximation is shown as the blue dashed line. Consistently with the one-dimensional model simulations and the perturbation analysis, the finite dc-like current remains even after the irradiation of linearly polarized light when the mean-field potential, or the Kohn--Sham potential, has the time dependence with the ALDA. By contrast, the unphysical constant current after the field irradiation is removed when the time-dependence of the mean-field is ignored in the independent particle approximation.

As seen from Fig.~\ref{fig:shift_current_tddft}, one can confirm that the unphysical dc-like due to the mean-field approximation significantly affects the photovoltaic response even in a realistic material, BaTiO$_3$, beyond the one-dimensional model calculations. Therefore, the investigation on the shift-current and injection-current with mean-field theories has to be carefully conducted by properly removing the unphysical dc-current response.

\section{Summary \label{sec:summary}} 

In this study, we theoretically investigated second-order nonlinear optical effects in solids, specifically focusing on the shift and the injection currents. We utilized numerical simulations based on mean-field theories, such as TDHF and TDH methods, to explore real-time electron dynamics under laser pulse excitation and the resulting nonlinear current in the time domain. The numerical simulations demonstrated that the dc-component of the second-order nonlinear current remains finite even after irradiation with linearly polarized light, indicating a possible induction of the injection current by the time-dependent mean field. However, simulations based on the independent particle approximation did not exhibit such a residual dc-component after linearly polarized light irradiation.

To understand the origin of this residual dc-component observed in the nonlinear current, we performed perturbation analysis using various levels of theories, including the exact many-body Schr\"odinger equation, mean-field approximations, and the independent particle approximation. The perturbation analysis with the exact many-body Schr\"odinger equation revealed that linearly polarized light cannot induce such residual dc-component in the second-order nonlinear current when the system has time-reversal symmetry before laser irradiation. Similarly, the perturbation analysis using the independent particle approximation arrived at the same conclusion, indicating that the presence of a residual dc-component in the second-order nonlinear current after laser irradiation may occur only if the applied field is elliptically polarized or breaks time-reversal symmetry. Consequently, the perturbation analysis clarified that the residual dc-component in the nonlinear current is an artifact of the mean-field approximations.

Further we performed perturbation analysis within the mean-field theories, revealing that the unphysical dc-component in the nonlinear current arises from population imbalances in the Brillouin zone, specifically at $\vecb k$ and $-\vecb k$ points. Additionally, the perturbation analysis showed that this unphysical population imbalance is caused by quantum interference between different excitation paths involving self-excitation paths opened via the time-dependent mean field that are not present in the full many-body solution. Furthermore, we performed the first-principles electron dynamics calculations based on TDDFT using the adiabatic local density approximation and confirmed that these findings apply also to realistic materials. The resulting residual current obtained by using the adiabatic approximation indicates a significance of a time-nonlocal memory effect in the TDDFT to capture the proper nonlinear dynamics of interacting many-body systems.

The unphysical dc current, resembling the injection current, induced by mean-field approximations may overcome the intrinsic shift-current contribution due to the higher susceptibility divergence of injection current compared to that of the shift current.

From a different perspective, the induction of unphysical current through self-excitation paths via mean fields suggests opportunities for improving density and current-based many-body theories to describe light-induced nonlinear phenomena more accurately. An accurate theory should prevent the induction of unphysical dc current by eliminating population imbalances arising from self-excitation paths. This could be achieved by properly designing the Hartree-exchange-correlation kernel, $f_{\mathrm{Hxc}}(\vecb r, \vecb r', \omega)$, in time-dependent density functional theory. While this study focused on second-order nonlinear optical responses, it would be crucial to consider the potential significant impact of self-excitation paths in even higher-order nonlinear phenomena as well. Therefore, the limitations of local and semi-local adiabatic approximations and the effects of unphysical self-excitation paths should be carefully evaluated for further investigations on highly-nonlinear optical phenomena.

\begin{acknowledgments}
This work was supported by JSPS KAKENHI Grant Numbers JP20K14382 and JP21H01842, the Cluster of Excellence 'CUI: Advanced Imaging of Matter'- EXC 2056 - project ID 390715994, SFB-925 "Light induced dynamics and control of correlated quantum systems" – project 170620586  of the Deutsche Forschungsgemeinschaft (DFG),  and the Max Planck-New York City Center for Non-Equilibrium Quantum Phenomena. The Flatiron Institute is a division of the Simons Foundation.
\end{acknowledgments}

\bibliography{ref}

\begin{thebibliography}{35}%
\makeatletter
\providecommand \@ifxundefined [1]{%
 \@ifx{#1\undefined}
}%
\providecommand \@ifnum [1]{%
 \ifnum #1\expandafter \@firstoftwo
 \else \expandafter \@secondoftwo
 \fi
}%
\providecommand \@ifx [1]{%
 \ifx #1\expandafter \@firstoftwo
 \else \expandafter \@secondoftwo
 \fi
}%
\providecommand \natexlab [1]{#1}%
\providecommand \enquote  [1]{``#1''}%
\providecommand \bibnamefont  [1]{#1}%
\providecommand \bibfnamefont [1]{#1}%
\providecommand \citenamefont [1]{#1}%
\providecommand \href@noop [0]{\@secondoftwo}%
\providecommand \href [0]{\begingroup \@sanitize@url \@href}%
\providecommand \@href[1]{\@@startlink{#1}\@@href}%
\providecommand \@@href[1]{\endgroup#1\@@endlink}%
\providecommand \@sanitize@url [0]{\catcode `\\12\catcode `\$12\catcode
  `\&12\catcode `\#12\catcode `\^12\catcode `\_12\catcode `\%12\relax}%
\providecommand \@@startlink[1]{}%
\providecommand \@@endlink[0]{}%
\providecommand \url  [0]{\begingroup\@sanitize@url \@url }%
\providecommand \@url [1]{\endgroup\@href {#1}{\urlprefix }}%
\providecommand \urlprefix  [0]{URL }%
\providecommand \Eprint [0]{\href }%
\providecommand \doibase [0]{http://dx.doi.org/}%
\providecommand \selectlanguage [0]{\@gobble}%
\providecommand \bibinfo  [0]{\@secondoftwo}%
\providecommand \bibfield  [0]{\@secondoftwo}%
\providecommand \translation [1]{[#1]}%
\providecommand \BibitemOpen [0]{}%
\providecommand \bibitemStop [0]{}%
\providecommand \bibitemNoStop [0]{.\EOS\space}%
\providecommand \EOS [0]{\spacefactor3000\relax}%
\providecommand \BibitemShut  [1]{\csname bibitem#1\endcsname}%
\let\auto@bib@innerbib\@empty
\bibitem [{\citenamefont {Seidel}\ \emph {et~al.}(2011)\citenamefont {Seidel},
  \citenamefont {Fu}, \citenamefont {Yang}, \citenamefont {Alarc\'on-Llad\'o},
  \citenamefont {Wu}, \citenamefont {Ramesh},\ and\ \citenamefont
  {Ager}}]{PhysRevLett.107.126805}%
  \BibitemOpen
  \bibfield  {author} {\bibinfo {author} {\bibfnamefont {J.}~\bibnamefont
  {Seidel}}, \bibinfo {author} {\bibfnamefont {D.}~\bibnamefont {Fu}}, \bibinfo
  {author} {\bibfnamefont {S.-Y.}\ \bibnamefont {Yang}}, \bibinfo {author}
  {\bibfnamefont {E.}~\bibnamefont {Alarc\'on-Llad\'o}}, \bibinfo {author}
  {\bibfnamefont {J.}~\bibnamefont {Wu}}, \bibinfo {author} {\bibfnamefont
  {R.}~\bibnamefont {Ramesh}}, \ and\ \bibinfo {author} {\bibfnamefont {J.~W.}\
  \bibnamefont {Ager}},\ }\href {\doibase 10.1103/PhysRevLett.107.126805}
  {\bibfield  {journal} {\bibinfo  {journal} {Phys. Rev. Lett.}\ }\textbf
  {\bibinfo {volume} {107}},\ \bibinfo {pages} {126805} (\bibinfo {year}
  {2011})}\BibitemShut {NoStop}%
\bibitem [{\citenamefont {Choi}\ \emph {et~al.}(2009)\citenamefont {Choi},
  \citenamefont {Lee}, \citenamefont {Choi}, \citenamefont {Kiryukhin},\ and\
  \citenamefont {Cheong}}]{doi:10.1126/science.1168636}%
  \BibitemOpen
  \bibfield  {author} {\bibinfo {author} {\bibfnamefont {T.}~\bibnamefont
  {Choi}}, \bibinfo {author} {\bibfnamefont {S.}~\bibnamefont {Lee}}, \bibinfo
  {author} {\bibfnamefont {Y.~J.}\ \bibnamefont {Choi}}, \bibinfo {author}
  {\bibfnamefont {V.}~\bibnamefont {Kiryukhin}}, \ and\ \bibinfo {author}
  {\bibfnamefont {S.-W.}\ \bibnamefont {Cheong}},\ }\href {\doibase
  10.1126/science.1168636} {\bibfield  {journal} {\bibinfo  {journal}
  {Science}\ }\textbf {\bibinfo {volume} {324}},\ \bibinfo {pages} {63}
  (\bibinfo {year} {2009})}\BibitemShut {NoStop}%
\bibitem [{\citenamefont {Yang}\ \emph {et~al.}(2010)\citenamefont {Yang},
  \citenamefont {Seidel}, \citenamefont {Byrnes}, \citenamefont {Shafer},
  \citenamefont {Yang}, \citenamefont {Rossell}, \citenamefont {Yu},
  \citenamefont {Chu}, \citenamefont {Scott}, \citenamefont {Ager},
  \citenamefont {Martin},\ and\ \citenamefont {Ramesh}}]{Yang2010}%
  \BibitemOpen
  \bibfield  {author} {\bibinfo {author} {\bibfnamefont {S.~Y.}\ \bibnamefont
  {Yang}}, \bibinfo {author} {\bibfnamefont {J.}~\bibnamefont {Seidel}},
  \bibinfo {author} {\bibfnamefont {S.~J.}\ \bibnamefont {Byrnes}}, \bibinfo
  {author} {\bibfnamefont {P.}~\bibnamefont {Shafer}}, \bibinfo {author}
  {\bibfnamefont {C.-H.}\ \bibnamefont {Yang}}, \bibinfo {author}
  {\bibfnamefont {M.~D.}\ \bibnamefont {Rossell}}, \bibinfo {author}
  {\bibfnamefont {P.}~\bibnamefont {Yu}}, \bibinfo {author} {\bibfnamefont
  {Y.-H.}\ \bibnamefont {Chu}}, \bibinfo {author} {\bibfnamefont {J.~F.}\
  \bibnamefont {Scott}}, \bibinfo {author} {\bibfnamefont {J.~W.}\ \bibnamefont
  {Ager}}, \bibinfo {author} {\bibfnamefont {L.~W.}\ \bibnamefont {Martin}}, \
  and\ \bibinfo {author} {\bibfnamefont {R.}~\bibnamefont {Ramesh}},\ }\href
  {\doibase 10.1038/nnano.2009.451} {\bibfield  {journal} {\bibinfo  {journal}
  {Nature Nanotechnology}\ }\textbf {\bibinfo {volume} {5}},\ \bibinfo {pages}
  {143} (\bibinfo {year} {2010})}\BibitemShut {NoStop}%
\bibitem [{\citenamefont {Koch}\ \emph {et~al.}(1975)\citenamefont {Koch},
  \citenamefont {Munser}, \citenamefont {Ruppel},\ and\ \citenamefont
  {Würfel}}]{KOCH1975847}%
  \BibitemOpen
  \bibfield  {author} {\bibinfo {author} {\bibfnamefont {W.}~\bibnamefont
  {Koch}}, \bibinfo {author} {\bibfnamefont {R.}~\bibnamefont {Munser}},
  \bibinfo {author} {\bibfnamefont {W.}~\bibnamefont {Ruppel}}, \ and\ \bibinfo
  {author} {\bibfnamefont {P.}~\bibnamefont {Würfel}},\ }\href {\doibase
  https://doi.org/10.1016/0038-1098(75)90735-8} {\bibfield  {journal} {\bibinfo
   {journal} {Solid State Communications}\ }\textbf {\bibinfo {volume} {17}},\
  \bibinfo {pages} {847} (\bibinfo {year} {1975})}\BibitemShut {NoStop}%
\bibitem [{\citenamefont {Fridkin}\ \emph {et~al.}(1974)\citenamefont
  {Fridkin}, \citenamefont {Grekov}, \citenamefont {Ionov}, \citenamefont
  {Rodin}, \citenamefont {Savchenko},\ and\ \citenamefont
  {Mikhailina}}]{doi:10.1080/00150197408234118}%
  \BibitemOpen
  \bibfield  {author} {\bibinfo {author} {\bibfnamefont {V.~M.}\ \bibnamefont
  {Fridkin}}, \bibinfo {author} {\bibfnamefont {A.~A.}\ \bibnamefont {Grekov}},
  \bibinfo {author} {\bibfnamefont {P.~V.}\ \bibnamefont {Ionov}}, \bibinfo
  {author} {\bibfnamefont {A.~I.}\ \bibnamefont {Rodin}}, \bibinfo {author}
  {\bibfnamefont {E.~A.}\ \bibnamefont {Savchenko}}, \ and\ \bibinfo {author}
  {\bibfnamefont {K.~A.}\ \bibnamefont {Mikhailina}},\ }\href {\doibase
  10.1080/00150197408234118} {\bibfield  {journal} {\bibinfo  {journal}
  {Ferroelectrics}\ }\textbf {\bibinfo {volume} {8}},\ \bibinfo {pages} {433}
  (\bibinfo {year} {1974})}\BibitemShut {NoStop}%
\bibitem [{\citenamefont {Krätzig}\ and\ \citenamefont
  {Kurz}(1976)}]{doi:10.1080/00150197608236593}%
  \BibitemOpen
  \bibfield  {author} {\bibinfo {author} {\bibfnamefont {E.}~\bibnamefont
  {Krätzig}}\ and\ \bibinfo {author} {\bibfnamefont {H.}~\bibnamefont
  {Kurz}},\ }\href {\doibase 10.1080/00150197608236593} {\bibfield  {journal}
  {\bibinfo  {journal} {Ferroelectrics}\ }\textbf {\bibinfo {volume} {13}},\
  \bibinfo {pages} {295} (\bibinfo {year} {1976})}\BibitemShut {NoStop}%
\bibitem [{\citenamefont {Krätzig}\ and\ \citenamefont
  {Kurz}(1977{\natexlab{a}})}]{Kratzig_1977}%
  \BibitemOpen
  \bibfield  {author} {\bibinfo {author} {\bibfnamefont {E.}~\bibnamefont
  {Krätzig}}\ and\ \bibinfo {author} {\bibfnamefont {H.}~\bibnamefont
  {Kurz}},\ }\href {\doibase 10.1149/1.2133226} {\bibfield  {journal} {\bibinfo
   {journal} {Journal of The Electrochemical Society}\ }\textbf {\bibinfo
  {volume} {124}},\ \bibinfo {pages} {131} (\bibinfo {year}
  {1977}{\natexlab{a}})}\BibitemShut {NoStop}%
\bibitem [{\citenamefont {Krätzig}\ and\ \citenamefont
  {Kurz}(1977{\natexlab{b}})}]{doi:10.1080/713819567}%
  \BibitemOpen
  \bibfield  {author} {\bibinfo {author} {\bibfnamefont {E.}~\bibnamefont
  {Krätzig}}\ and\ \bibinfo {author} {\bibfnamefont {H.}~\bibnamefont
  {Kurz}},\ }\href {\doibase 10.1080/713819567} {\bibfield  {journal} {\bibinfo
   {journal} {Optica Acta: International Journal of Optics}\ }\textbf {\bibinfo
  {volume} {24}},\ \bibinfo {pages} {475} (\bibinfo {year}
  {1977}{\natexlab{b}})}\BibitemShut {NoStop}%
\bibitem [{\citenamefont {Koch}\ \emph {et~al.}(1976)\citenamefont {Koch},
  \citenamefont {Munser}, \citenamefont {Ruppel},\ and\ \citenamefont
  {Würfel}}]{doi:10.1080/00150197608236596}%
  \BibitemOpen
  \bibfield  {author} {\bibinfo {author} {\bibfnamefont {W.~T.~H.}\
  \bibnamefont {Koch}}, \bibinfo {author} {\bibfnamefont {R.}~\bibnamefont
  {Munser}}, \bibinfo {author} {\bibfnamefont {W.}~\bibnamefont {Ruppel}}, \
  and\ \bibinfo {author} {\bibfnamefont {P.}~\bibnamefont {Würfel}},\ }\href
  {\doibase 10.1080/00150197608236596} {\bibfield  {journal} {\bibinfo
  {journal} {Ferroelectrics}\ }\textbf {\bibinfo {volume} {13}},\ \bibinfo
  {pages} {305} (\bibinfo {year} {1976})}\BibitemShut {NoStop}%
\bibitem [{\citenamefont {Sipe}\ and\ \citenamefont
  {Shkrebtii}(2000)}]{PhysRevB.61.5337}%
  \BibitemOpen
  \bibfield  {author} {\bibinfo {author} {\bibfnamefont {J.~E.}\ \bibnamefont
  {Sipe}}\ and\ \bibinfo {author} {\bibfnamefont {A.~I.}\ \bibnamefont
  {Shkrebtii}},\ }\href {\doibase 10.1103/PhysRevB.61.5337} {\bibfield
  {journal} {\bibinfo  {journal} {Phys. Rev. B}\ }\textbf {\bibinfo {volume}
  {61}},\ \bibinfo {pages} {5337} (\bibinfo {year} {2000})}\BibitemShut
  {NoStop}%
\bibitem [{\citenamefont {Young}\ and\ \citenamefont
  {Rappe}(2012)}]{PhysRevLett.109.116601}%
  \BibitemOpen
  \bibfield  {author} {\bibinfo {author} {\bibfnamefont {S.~M.}\ \bibnamefont
  {Young}}\ and\ \bibinfo {author} {\bibfnamefont {A.~M.}\ \bibnamefont
  {Rappe}},\ }\href {\doibase 10.1103/PhysRevLett.109.116601} {\bibfield
  {journal} {\bibinfo  {journal} {Phys. Rev. Lett.}\ }\textbf {\bibinfo
  {volume} {109}},\ \bibinfo {pages} {116601} (\bibinfo {year}
  {2012})}\BibitemShut {NoStop}%
\bibitem [{\citenamefont {Young}\ \emph {et~al.}(2012)\citenamefont {Young},
  \citenamefont {Zheng},\ and\ \citenamefont {Rappe}}]{PhysRevLett.109.236601}%
  \BibitemOpen
  \bibfield  {author} {\bibinfo {author} {\bibfnamefont {S.~M.}\ \bibnamefont
  {Young}}, \bibinfo {author} {\bibfnamefont {F.}~\bibnamefont {Zheng}}, \ and\
  \bibinfo {author} {\bibfnamefont {A.~M.}\ \bibnamefont {Rappe}},\ }\href
  {\doibase 10.1103/PhysRevLett.109.236601} {\bibfield  {journal} {\bibinfo
  {journal} {Phys. Rev. Lett.}\ }\textbf {\bibinfo {volume} {109}},\ \bibinfo
  {pages} {236601} (\bibinfo {year} {2012})}\BibitemShut {NoStop}%
\bibitem [{\citenamefont {Cook}\ \emph {et~al.}(2017)\citenamefont {Cook},
  \citenamefont {M.~Fregoso}, \citenamefont {de~Juan}, \citenamefont {Coh},\
  and\ \citenamefont {Moore}}]{Cook2017}%
  \BibitemOpen
  \bibfield  {author} {\bibinfo {author} {\bibfnamefont {A.~M.}\ \bibnamefont
  {Cook}}, \bibinfo {author} {\bibfnamefont {B.}~\bibnamefont {M.~Fregoso}},
  \bibinfo {author} {\bibfnamefont {F.}~\bibnamefont {de~Juan}}, \bibinfo
  {author} {\bibfnamefont {S.}~\bibnamefont {Coh}}, \ and\ \bibinfo {author}
  {\bibfnamefont {J.~E.}\ \bibnamefont {Moore}},\ }\href {\doibase
  10.1038/ncomms14176} {\bibfield  {journal} {\bibinfo  {journal} {Nature
  Communications}\ }\textbf {\bibinfo {volume} {8}},\ \bibinfo {pages} {14176}
  (\bibinfo {year} {2017})}\BibitemShut {NoStop}%
\bibitem [{\citenamefont {Zhang}\ \emph {et~al.}(2018)\citenamefont {Zhang},
  \citenamefont {Ishizuka}, \citenamefont {van~den Brink}, \citenamefont
  {Felser}, \citenamefont {Yan},\ and\ \citenamefont
  {Nagaosa}}]{PhysRevB.97.241118}%
  \BibitemOpen
  \bibfield  {author} {\bibinfo {author} {\bibfnamefont {Y.}~\bibnamefont
  {Zhang}}, \bibinfo {author} {\bibfnamefont {H.}~\bibnamefont {Ishizuka}},
  \bibinfo {author} {\bibfnamefont {J.}~\bibnamefont {van~den Brink}}, \bibinfo
  {author} {\bibfnamefont {C.}~\bibnamefont {Felser}}, \bibinfo {author}
  {\bibfnamefont {B.}~\bibnamefont {Yan}}, \ and\ \bibinfo {author}
  {\bibfnamefont {N.}~\bibnamefont {Nagaosa}},\ }\href {\doibase
  10.1103/PhysRevB.97.241118} {\bibfield  {journal} {\bibinfo  {journal} {Phys.
  Rev. B}\ }\textbf {\bibinfo {volume} {97}},\ \bibinfo {pages} {241118}
  (\bibinfo {year} {2018})}\BibitemShut {NoStop}%
\bibitem [{\citenamefont {Fei}\ \emph {et~al.}(2020)\citenamefont {Fei},
  \citenamefont {Tan},\ and\ \citenamefont {Rappe}}]{PhysRevB.101.045104}%
  \BibitemOpen
  \bibfield  {author} {\bibinfo {author} {\bibfnamefont {R.}~\bibnamefont
  {Fei}}, \bibinfo {author} {\bibfnamefont {L.~Z.}\ \bibnamefont {Tan}}, \ and\
  \bibinfo {author} {\bibfnamefont {A.~M.}\ \bibnamefont {Rappe}},\ }\href
  {\doibase 10.1103/PhysRevB.101.045104} {\bibfield  {journal} {\bibinfo
  {journal} {Phys. Rev. B}\ }\textbf {\bibinfo {volume} {101}},\ \bibinfo
  {pages} {045104} (\bibinfo {year} {2020})}\BibitemShut {NoStop}%
\bibitem [{\citenamefont {Laman}\ \emph {et~al.}(2005)\citenamefont {Laman},
  \citenamefont {Bieler},\ and\ \citenamefont {van Driel}}]{10.1063/1.2131191}%
  \BibitemOpen
  \bibfield  {author} {\bibinfo {author} {\bibfnamefont {N.}~\bibnamefont
  {Laman}}, \bibinfo {author} {\bibfnamefont {M.}~\bibnamefont {Bieler}}, \
  and\ \bibinfo {author} {\bibfnamefont {H.~M.}\ \bibnamefont {van Driel}},\
  }\href {\doibase 10.1063/1.2131191} {\bibfield  {journal} {\bibinfo
  {journal} {Journal of Applied Physics}\ }\textbf {\bibinfo {volume} {98}},\
  \bibinfo {pages} {103507} (\bibinfo {year} {2005})}\BibitemShut {NoStop}%
\bibitem [{\citenamefont {Ogawa}\ \emph {et~al.}(2017)\citenamefont {Ogawa},
  \citenamefont {Sotome}, \citenamefont {Kaneko}, \citenamefont {Ogino},\ and\
  \citenamefont {Tokura}}]{PhysRevB.96.241203}%
  \BibitemOpen
  \bibfield  {author} {\bibinfo {author} {\bibfnamefont {N.}~\bibnamefont
  {Ogawa}}, \bibinfo {author} {\bibfnamefont {M.}~\bibnamefont {Sotome}},
  \bibinfo {author} {\bibfnamefont {Y.}~\bibnamefont {Kaneko}}, \bibinfo
  {author} {\bibfnamefont {M.}~\bibnamefont {Ogino}}, \ and\ \bibinfo {author}
  {\bibfnamefont {Y.}~\bibnamefont {Tokura}},\ }\href {\doibase
  10.1103/PhysRevB.96.241203} {\bibfield  {journal} {\bibinfo  {journal} {Phys.
  Rev. B}\ }\textbf {\bibinfo {volume} {96}},\ \bibinfo {pages} {241203}
  (\bibinfo {year} {2017})}\BibitemShut {NoStop}%
\bibitem [{\citenamefont {Nakamura}\ \emph {et~al.}(2018)\citenamefont
  {Nakamura}, \citenamefont {Hatada}, \citenamefont {Kaneko}, \citenamefont
  {Ogawa}, \citenamefont {Tokura},\ and\ \citenamefont
  {Kawasaki}}]{10.1063/1.5055692}%
  \BibitemOpen
  \bibfield  {author} {\bibinfo {author} {\bibfnamefont {M.}~\bibnamefont
  {Nakamura}}, \bibinfo {author} {\bibfnamefont {H.}~\bibnamefont {Hatada}},
  \bibinfo {author} {\bibfnamefont {Y.}~\bibnamefont {Kaneko}}, \bibinfo
  {author} {\bibfnamefont {N.}~\bibnamefont {Ogawa}}, \bibinfo {author}
  {\bibfnamefont {Y.}~\bibnamefont {Tokura}}, \ and\ \bibinfo {author}
  {\bibfnamefont {M.}~\bibnamefont {Kawasaki}},\ }\href {\doibase
  10.1063/1.5055692} {\bibfield  {journal} {\bibinfo  {journal} {Applied
  Physics Letters}\ }\textbf {\bibinfo {volume} {113}},\ \bibinfo {pages}
  {232901} (\bibinfo {year} {2018})}\BibitemShut {NoStop}%
\bibitem [{\citenamefont {Nakamura}\ \emph {et~al.}(2017)\citenamefont
  {Nakamura}, \citenamefont {Horiuchi}, \citenamefont {Kagawa}, \citenamefont
  {Ogawa}, \citenamefont {Kurumaji}, \citenamefont {Tokura},\ and\
  \citenamefont {Kawasaki}}]{Nakamura2017}%
  \BibitemOpen
  \bibfield  {author} {\bibinfo {author} {\bibfnamefont {M.}~\bibnamefont
  {Nakamura}}, \bibinfo {author} {\bibfnamefont {S.}~\bibnamefont {Horiuchi}},
  \bibinfo {author} {\bibfnamefont {F.}~\bibnamefont {Kagawa}}, \bibinfo
  {author} {\bibfnamefont {N.}~\bibnamefont {Ogawa}}, \bibinfo {author}
  {\bibfnamefont {T.}~\bibnamefont {Kurumaji}}, \bibinfo {author}
  {\bibfnamefont {Y.}~\bibnamefont {Tokura}}, \ and\ \bibinfo {author}
  {\bibfnamefont {M.}~\bibnamefont {Kawasaki}},\ }\href {\doibase
  10.1038/s41467-017-00250-y} {\bibfield  {journal} {\bibinfo  {journal}
  {Nature Communications}\ }\textbf {\bibinfo {volume} {8}},\ \bibinfo {pages}
  {281} (\bibinfo {year} {2017})}\BibitemShut {NoStop}%
\bibitem [{\citenamefont {Sotome}\ \emph
  {et~al.}(2019{\natexlab{a}})\citenamefont {Sotome}, \citenamefont {Nakamura},
  \citenamefont {Fujioka}, \citenamefont {Ogino}, \citenamefont {Kaneko},
  \citenamefont {Morimoto}, \citenamefont {Zhang}, \citenamefont {Kawasaki},
  \citenamefont {Nagaosa}, \citenamefont {Tokura},\ and\ \citenamefont
  {Ogawa}}]{doi:10.1073/pnas.1802427116}%
  \BibitemOpen
  \bibfield  {author} {\bibinfo {author} {\bibfnamefont {M.}~\bibnamefont
  {Sotome}}, \bibinfo {author} {\bibfnamefont {M.}~\bibnamefont {Nakamura}},
  \bibinfo {author} {\bibfnamefont {J.}~\bibnamefont {Fujioka}}, \bibinfo
  {author} {\bibfnamefont {M.}~\bibnamefont {Ogino}}, \bibinfo {author}
  {\bibfnamefont {Y.}~\bibnamefont {Kaneko}}, \bibinfo {author} {\bibfnamefont
  {T.}~\bibnamefont {Morimoto}}, \bibinfo {author} {\bibfnamefont
  {Y.}~\bibnamefont {Zhang}}, \bibinfo {author} {\bibfnamefont
  {M.}~\bibnamefont {Kawasaki}}, \bibinfo {author} {\bibfnamefont
  {N.}~\bibnamefont {Nagaosa}}, \bibinfo {author} {\bibfnamefont
  {Y.}~\bibnamefont {Tokura}}, \ and\ \bibinfo {author} {\bibfnamefont
  {N.}~\bibnamefont {Ogawa}},\ }\href {\doibase 10.1073/pnas.1802427116}
  {\bibfield  {journal} {\bibinfo  {journal} {Proceedings of the National
  Academy of Sciences}\ }\textbf {\bibinfo {volume} {116}},\ \bibinfo {pages}
  {1929} (\bibinfo {year} {2019}{\natexlab{a}})}\BibitemShut {NoStop}%
\bibitem [{\citenamefont {Sotome}\ \emph
  {et~al.}(2019{\natexlab{b}})\citenamefont {Sotome}, \citenamefont {Nakamura},
  \citenamefont {Fujioka}, \citenamefont {Ogino}, \citenamefont {Kaneko},
  \citenamefont {Morimoto}, \citenamefont {Zhang}, \citenamefont {Kawasaki},
  \citenamefont {Nagaosa}, \citenamefont {Tokura},\ and\ \citenamefont
  {Ogawa}}]{10.1063/1.5087960}%
  \BibitemOpen
  \bibfield  {author} {\bibinfo {author} {\bibfnamefont {M.}~\bibnamefont
  {Sotome}}, \bibinfo {author} {\bibfnamefont {M.}~\bibnamefont {Nakamura}},
  \bibinfo {author} {\bibfnamefont {J.}~\bibnamefont {Fujioka}}, \bibinfo
  {author} {\bibfnamefont {M.}~\bibnamefont {Ogino}}, \bibinfo {author}
  {\bibfnamefont {Y.}~\bibnamefont {Kaneko}}, \bibinfo {author} {\bibfnamefont
  {T.}~\bibnamefont {Morimoto}}, \bibinfo {author} {\bibfnamefont
  {Y.}~\bibnamefont {Zhang}}, \bibinfo {author} {\bibfnamefont
  {M.}~\bibnamefont {Kawasaki}}, \bibinfo {author} {\bibfnamefont
  {N.}~\bibnamefont {Nagaosa}}, \bibinfo {author} {\bibfnamefont
  {Y.}~\bibnamefont {Tokura}}, \ and\ \bibinfo {author} {\bibfnamefont
  {N.}~\bibnamefont {Ogawa}},\ }\href {\doibase 10.1063/1.5087960} {\bibfield
  {journal} {\bibinfo  {journal} {Applied Physics Letters}\ }\textbf {\bibinfo
  {volume} {114}},\ \bibinfo {pages} {151101} (\bibinfo {year}
  {2019}{\natexlab{b}})}\BibitemShut {NoStop}%
\bibitem [{\citenamefont {Chan}\ \emph {et~al.}(2021)\citenamefont {Chan},
  \citenamefont {Qiu}, \citenamefont {da~Jornada},\ and\ \citenamefont
  {Louie}}]{doi:10.1073/pnas.1906938118}%
  \BibitemOpen
  \bibfield  {author} {\bibinfo {author} {\bibfnamefont {Y.-H.}\ \bibnamefont
  {Chan}}, \bibinfo {author} {\bibfnamefont {D.~Y.}\ \bibnamefont {Qiu}},
  \bibinfo {author} {\bibfnamefont {F.~H.}\ \bibnamefont {da~Jornada}}, \ and\
  \bibinfo {author} {\bibfnamefont {S.~G.}\ \bibnamefont {Louie}},\ }\href
  {\doibase 10.1073/pnas.1906938118} {\bibfield  {journal} {\bibinfo  {journal}
  {Proceedings of the National Academy of Sciences}\ }\textbf {\bibinfo
  {volume} {118}},\ \bibinfo {pages} {e1906938118} (\bibinfo {year}
  {2021})}\BibitemShut {NoStop}%
\bibitem [{\citenamefont {Kaneko}\ \emph {et~al.}(2021)\citenamefont {Kaneko},
  \citenamefont {Sun}, \citenamefont {Murakami}, \citenamefont
  {Gole\ifmmode~\check{z}\else \v{z}\fi{}},\ and\ \citenamefont
  {Millis}}]{PhysRevLett.127.127402}%
  \BibitemOpen
  \bibfield  {author} {\bibinfo {author} {\bibfnamefont {T.}~\bibnamefont
  {Kaneko}}, \bibinfo {author} {\bibfnamefont {Z.}~\bibnamefont {Sun}},
  \bibinfo {author} {\bibfnamefont {Y.}~\bibnamefont {Murakami}}, \bibinfo
  {author} {\bibfnamefont {D.}~\bibnamefont {Gole\ifmmode~\check{z}\else
  \v{z}\fi{}}}, \ and\ \bibinfo {author} {\bibfnamefont {A.~J.}\ \bibnamefont
  {Millis}},\ }\href {\doibase 10.1103/PhysRevLett.127.127402} {\bibfield
  {journal} {\bibinfo  {journal} {Phys. Rev. Lett.}\ }\textbf {\bibinfo
  {volume} {127}},\ \bibinfo {pages} {127402} (\bibinfo {year}
  {2021})}\BibitemShut {NoStop}%
\bibitem [{\citenamefont {Kwei}\ \emph {et~al.}(1993)\citenamefont {Kwei},
  \citenamefont {Lawson}, \citenamefont {Billinge},\ and\ \citenamefont
  {Cheong}}]{Kwei1993}%
  \BibitemOpen
  \bibfield  {author} {\bibinfo {author} {\bibfnamefont {G.~H.}\ \bibnamefont
  {Kwei}}, \bibinfo {author} {\bibfnamefont {A.~C.}\ \bibnamefont {Lawson}},
  \bibinfo {author} {\bibfnamefont {S.~J.~L.}\ \bibnamefont {Billinge}}, \ and\
  \bibinfo {author} {\bibfnamefont {S.~W.}\ \bibnamefont {Cheong}},\ }\href
  {\doibase 10.1021/j100112a043} {\bibfield  {journal} {\bibinfo  {journal}
  {The Journal of Physical Chemistry}\ }\textbf {\bibinfo {volume} {97}},\
  \bibinfo {pages} {2368} (\bibinfo {year} {1993})}\BibitemShut {NoStop}%
\bibitem [{\citenamefont {Onida}\ \emph {et~al.}(2002)\citenamefont {Onida},
  \citenamefont {Reining},\ and\ \citenamefont {Rubio}}]{RevModPhys.74.601}%
  \BibitemOpen
  \bibfield  {author} {\bibinfo {author} {\bibfnamefont {G.}~\bibnamefont
  {Onida}}, \bibinfo {author} {\bibfnamefont {L.}~\bibnamefont {Reining}}, \
  and\ \bibinfo {author} {\bibfnamefont {A.}~\bibnamefont {Rubio}},\ }\href
  {\doibase 10.1103/RevModPhys.74.601} {\bibfield  {journal} {\bibinfo
  {journal} {Rev. Mod. Phys.}\ }\textbf {\bibinfo {volume} {74}},\ \bibinfo
  {pages} {601} (\bibinfo {year} {2002})}\BibitemShut {NoStop}%
\bibitem [{\citenamefont {Boyd}(2020)}]{boyd2020nonlinear}%
  \BibitemOpen
  \bibfield  {author} {\bibinfo {author} {\bibfnamefont {R.~W.}\ \bibnamefont
  {Boyd}},\ }\href@noop {} {\emph {\bibinfo {title} {Nonlinear optics}}}\
  (\bibinfo  {publisher} {Academic press},\ \bibinfo {year} {2020})\BibitemShut
  {NoStop}%
\bibitem [{\citenamefont {Dai}\ and\ \citenamefont
  {Rappe}(2023)}]{10.1063/5.0101513}%
  \BibitemOpen
  \bibfield  {author} {\bibinfo {author} {\bibfnamefont {Z.}~\bibnamefont
  {Dai}}\ and\ \bibinfo {author} {\bibfnamefont {A.~M.}\ \bibnamefont
  {Rappe}},\ }\href {\doibase 10.1063/5.0101513} {\bibfield  {journal}
  {\bibinfo  {journal} {Chemical Physics Reviews}\ }\textbf {\bibinfo {volume}
  {4}},\ \bibinfo {pages} {011303} (\bibinfo {year} {2023})}\BibitemShut
  {NoStop}%
\bibitem [{\citenamefont {Ghahramani}\ \emph {et~al.}(1991)\citenamefont
  {Ghahramani}, \citenamefont {Moss},\ and\ \citenamefont
  {Sipe}}]{PhysRevB.43.8990}%
  \BibitemOpen
  \bibfield  {author} {\bibinfo {author} {\bibfnamefont {E.}~\bibnamefont
  {Ghahramani}}, \bibinfo {author} {\bibfnamefont {D.~J.}\ \bibnamefont
  {Moss}}, \ and\ \bibinfo {author} {\bibfnamefont {J.~E.}\ \bibnamefont
  {Sipe}},\ }\href {\doibase 10.1103/PhysRevB.43.8990} {\bibfield  {journal}
  {\bibinfo  {journal} {Phys. Rev. B}\ }\textbf {\bibinfo {volume} {43}},\
  \bibinfo {pages} {8990} (\bibinfo {year} {1991})}\BibitemShut {NoStop}%
\bibitem [{\citenamefont {Aversa}\ and\ \citenamefont
  {Sipe}(1995)}]{PhysRevB.52.14636}%
  \BibitemOpen
  \bibfield  {author} {\bibinfo {author} {\bibfnamefont {C.}~\bibnamefont
  {Aversa}}\ and\ \bibinfo {author} {\bibfnamefont {J.~E.}\ \bibnamefont
  {Sipe}},\ }\href {\doibase 10.1103/PhysRevB.52.14636} {\bibfield  {journal}
  {\bibinfo  {journal} {Phys. Rev. B}\ }\textbf {\bibinfo {volume} {52}},\
  \bibinfo {pages} {14636} (\bibinfo {year} {1995})}\BibitemShut {NoStop}%
\bibitem [{\citenamefont {LANGHOFF}\ \emph {et~al.}(1972)\citenamefont
  {LANGHOFF}, \citenamefont {EPSTEIN},\ and\ \citenamefont
  {KARPLUS}}]{RevModPhys.44.602}%
  \BibitemOpen
  \bibfield  {author} {\bibinfo {author} {\bibfnamefont {P.~W.}\ \bibnamefont
  {LANGHOFF}}, \bibinfo {author} {\bibfnamefont {S.~T.}\ \bibnamefont
  {EPSTEIN}}, \ and\ \bibinfo {author} {\bibfnamefont {M.}~\bibnamefont
  {KARPLUS}},\ }\href {\doibase 10.1103/RevModPhys.44.602} {\bibfield
  {journal} {\bibinfo  {journal} {Rev. Mod. Phys.}\ }\textbf {\bibinfo {volume}
  {44}},\ \bibinfo {pages} {602} (\bibinfo {year} {1972})}\BibitemShut
  {NoStop}%
\bibitem [{\citenamefont {Perdew}\ and\ \citenamefont
  {Wang}(1992)}]{PhysRevB.45.13244}%
  \BibitemOpen
  \bibfield  {author} {\bibinfo {author} {\bibfnamefont {J.~P.}\ \bibnamefont
  {Perdew}}\ and\ \bibinfo {author} {\bibfnamefont {Y.}~\bibnamefont {Wang}},\
  }\href {\doibase 10.1103/PhysRevB.45.13244} {\bibfield  {journal} {\bibinfo
  {journal} {Phys. Rev. B}\ }\textbf {\bibinfo {volume} {45}},\ \bibinfo
  {pages} {13244} (\bibinfo {year} {1992})}\BibitemShut {NoStop}%
\bibitem [{\citenamefont {Reis}\ \emph {et~al.}(2003)\citenamefont {Reis},
  \citenamefont {Pacheco},\ and\ \citenamefont {Martins}}]{PhysRevB.68.155111}%
  \BibitemOpen
  \bibfield  {author} {\bibinfo {author} {\bibfnamefont {C.~L.}\ \bibnamefont
  {Reis}}, \bibinfo {author} {\bibfnamefont {J.~M.}\ \bibnamefont {Pacheco}}, \
  and\ \bibinfo {author} {\bibfnamefont {J.~L.}\ \bibnamefont {Martins}},\
  }\href {\doibase 10.1103/PhysRevB.68.155111} {\bibfield  {journal} {\bibinfo
  {journal} {Phys. Rev. B}\ }\textbf {\bibinfo {volume} {68}},\ \bibinfo
  {pages} {155111} (\bibinfo {year} {2003})}\BibitemShut {NoStop}%
\bibitem [{\citenamefont {Oliveira}\ and\ \citenamefont
  {Nogueira}(2008)}]{OLIVEIRA2008524}%
  \BibitemOpen
  \bibfield  {author} {\bibinfo {author} {\bibfnamefont {M.~J.}\ \bibnamefont
  {Oliveira}}\ and\ \bibinfo {author} {\bibfnamefont {F.}~\bibnamefont
  {Nogueira}},\ }\href {\doibase https://doi.org/10.1016/j.cpc.2007.11.003}
  {\bibfield  {journal} {\bibinfo  {journal} {Computer Physics Communications}\
  }\textbf {\bibinfo {volume} {178}},\ \bibinfo {pages} {524} (\bibinfo {year}
  {2008})}\BibitemShut {NoStop}%
\bibitem [{\citenamefont {Hartwigsen}\ \emph {et~al.}(1998)\citenamefont
  {Hartwigsen}, \citenamefont {Goedecker},\ and\ \citenamefont
  {Hutter}}]{PhysRevB.58.3641}%
  \BibitemOpen
  \bibfield  {author} {\bibinfo {author} {\bibfnamefont {C.}~\bibnamefont
  {Hartwigsen}}, \bibinfo {author} {\bibfnamefont {S.}~\bibnamefont
  {Goedecker}}, \ and\ \bibinfo {author} {\bibfnamefont {J.}~\bibnamefont
  {Hutter}},\ }\href {\doibase 10.1103/PhysRevB.58.3641} {\bibfield  {journal}
  {\bibinfo  {journal} {Phys. Rev. B}\ }\textbf {\bibinfo {volume} {58}},\
  \bibinfo {pages} {3641} (\bibinfo {year} {1998})}\BibitemShut {NoStop}%
\bibitem [{\citenamefont {Tancogne-Dejean}\ \emph {et~al.}(2020)\citenamefont
  {Tancogne-Dejean}, \citenamefont {Oliveira}, \citenamefont {Andrade},
  \citenamefont {Appel}, \citenamefont {Borca}, \citenamefont {Le~Breton},
  \citenamefont {Buchholz}, \citenamefont {Castro}, \citenamefont {Corni},
  \citenamefont {Correa}, \citenamefont {De~Giovannini}, \citenamefont
  {Delgado}, \citenamefont {Eich}, \citenamefont {Flick}, \citenamefont {Gil},
  \citenamefont {Gomez}, \citenamefont {Helbig}, \citenamefont {Hübener},
  \citenamefont {Jestädt}, \citenamefont {Jornet-Somoza}, \citenamefont
  {Larsen}, \citenamefont {Lebedeva}, \citenamefont {Lüders}, \citenamefont
  {Marques}, \citenamefont {Ohlmann}, \citenamefont {Pipolo}, \citenamefont
  {Rampp}, \citenamefont {Rozzi}, \citenamefont {Strubbe}, \citenamefont
  {Sato}, \citenamefont {Schäfer}, \citenamefont {Theophilou}, \citenamefont
  {Welden},\ and\ \citenamefont {Rubio}}]{10.1063/1.5142502}%
  \BibitemOpen
  \bibfield  {author} {\bibinfo {author} {\bibfnamefont {N.}~\bibnamefont
  {Tancogne-Dejean}}, \bibinfo {author} {\bibfnamefont {M.~J.~T.}\ \bibnamefont
  {Oliveira}}, \bibinfo {author} {\bibfnamefont {X.}~\bibnamefont {Andrade}},
  \bibinfo {author} {\bibfnamefont {H.}~\bibnamefont {Appel}}, \bibinfo
  {author} {\bibfnamefont {C.~H.}\ \bibnamefont {Borca}}, \bibinfo {author}
  {\bibfnamefont {G.}~\bibnamefont {Le~Breton}}, \bibinfo {author}
  {\bibfnamefont {F.}~\bibnamefont {Buchholz}}, \bibinfo {author}
  {\bibfnamefont {A.}~\bibnamefont {Castro}}, \bibinfo {author} {\bibfnamefont
  {S.}~\bibnamefont {Corni}}, \bibinfo {author} {\bibfnamefont {A.~A.}\
  \bibnamefont {Correa}}, \bibinfo {author} {\bibfnamefont {U.}~\bibnamefont
  {De~Giovannini}}, \bibinfo {author} {\bibfnamefont {A.}~\bibnamefont
  {Delgado}}, \bibinfo {author} {\bibfnamefont {F.~G.}\ \bibnamefont {Eich}},
  \bibinfo {author} {\bibfnamefont {J.}~\bibnamefont {Flick}}, \bibinfo
  {author} {\bibfnamefont {G.}~\bibnamefont {Gil}}, \bibinfo {author}
  {\bibfnamefont {A.}~\bibnamefont {Gomez}}, \bibinfo {author} {\bibfnamefont
  {N.}~\bibnamefont {Helbig}}, \bibinfo {author} {\bibfnamefont
  {H.}~\bibnamefont {Hübener}}, \bibinfo {author} {\bibfnamefont
  {R.}~\bibnamefont {Jestädt}}, \bibinfo {author} {\bibfnamefont
  {J.}~\bibnamefont {Jornet-Somoza}}, \bibinfo {author} {\bibfnamefont {A.~H.}\
  \bibnamefont {Larsen}}, \bibinfo {author} {\bibfnamefont {I.~V.}\
  \bibnamefont {Lebedeva}}, \bibinfo {author} {\bibfnamefont {M.}~\bibnamefont
  {Lüders}}, \bibinfo {author} {\bibfnamefont {M.~A.~L.}\ \bibnamefont
  {Marques}}, \bibinfo {author} {\bibfnamefont {S.~T.}\ \bibnamefont
  {Ohlmann}}, \bibinfo {author} {\bibfnamefont {S.}~\bibnamefont {Pipolo}},
  \bibinfo {author} {\bibfnamefont {M.}~\bibnamefont {Rampp}}, \bibinfo
  {author} {\bibfnamefont {C.~A.}\ \bibnamefont {Rozzi}}, \bibinfo {author}
  {\bibfnamefont {D.~A.}\ \bibnamefont {Strubbe}}, \bibinfo {author}
  {\bibfnamefont {S.~A.}\ \bibnamefont {Sato}}, \bibinfo {author}
  {\bibfnamefont {C.}~\bibnamefont {Schäfer}}, \bibinfo {author}
  {\bibfnamefont {I.}~\bibnamefont {Theophilou}}, \bibinfo {author}
  {\bibfnamefont {A.}~\bibnamefont {Welden}}, \ and\ \bibinfo {author}
  {\bibfnamefont {A.}~\bibnamefont {Rubio}},\ }\href {\doibase
  10.1063/1.5142502} {\bibfield  {journal} {\bibinfo  {journal} {The Journal of
  Chemical Physics}\ }\textbf {\bibinfo {volume} {152}},\ \bibinfo {pages}
  {124119} (\bibinfo {year} {2020})}\BibitemShut {NoStop}%
\end{thebibliography}%

\appendix

\end{document}